\documentclass[reprint, superscriptaddress, nofootinbib, longbibliography]{revtex4-1}

\usepackage{custom}
\usepackage{mathtools}

\usepackage{enumitem}

\usepackage{amsthm}
\usepackage{amsmath}
\def\be{\begin{equation}}
\def\ee{\end{equation}}
\def\ba{\begin{eqnarray}}
\def\ea{\end{eqnarray}}
\def\f{\frac}

\begin{document}
\title{Matter relative to quantum hypersurfaces}

\author{Philipp A. H\"{o}hn}
\email[]{philipp.hoehn@oist.jp}
\affiliation{Okinawa Institute of Science and Technology Graduate University, Onna, Okinawa 904 0495, Japan}
\affiliation{Department of Physics and Astronomy, University College London, London, United Kingdom}

\author{Andrea Russo}
\email[]{andrearusso.physics@gmail.com}
\affiliation{Department of Physics and Astronomy, University College London, London, United Kingdom}

\author{Alexander R. H. Smith}
\email[]{arhsmith@anselm.edu }
\affiliation{Department of Physics, Saint Anselm College, Manchester, New Hampshire 03102, USA} \affiliation{Department of Physics and Astronomy, Dartmouth College, Hanover, New Hampshire 03755, USA}

\date{\today}

\begin{abstract}
We explore the canonical description of a scalar field as a parameterized field theory on an extended phase space that includes additional embedding fields that characterize spacetime hypersurfaces $\mathsf{X}$ relative to which the scalar field is described. This theory is quantized via the Dirac prescription and physical states of the theory are used to define conditional wave functionals $\ket{\psi_\phi[\mathsf{X}]}$ interpreted as the state of the field relative to the hypersurface $\mathsf{X}$, thereby extending the Page-Wootters formalism to quantum field theory.
It is shown that this conditional wave functional satisfies the Tomonaga-Schwinger equation, thus demonstrating 
the formal equivalence between this extended Page-Wootters formalism and standard quantum field theory. We also construct relational Dirac observables and define a quantum deparameterization of the physical Hilbert space leading to a relational Heisenberg picture, which are both shown to be unitarily equivalent to the Page-Wootters formalism. Moreover, by treating hypersurfaces as quantum reference frames, we extend recently developed quantum frame transformations to changes between classical and nonclassical hypersurfaces. This allows us to exhibit the transformation properties of a quantum field under a larger class of transformations, which leads to a frame-dependent particle creation effect.
\end{abstract}

\maketitle


\section{Introduction}
\label{Sec_Introduction}

Physical theories that are independent of background spacetime structure are typically characterized by constraints. For example, in the absence of a boundary, the canonical Hamiltonian of general relativity vanishes on-shell.  In such theories, the Hamiltonian does not generate an evolution of the physical degrees of freedom; instead, dynamics emerges from relations between internal subsystems. As a consequence, a relational notion of dynamics has long been recognized as a conservative expectation of a quantum theory of gravity~\cite{kucharTimeInterpretationsQuantum1992,Isham1993,smolin_case_2006,Rovelli:1990ph,rovelliQuantumGravity2004,rovelli_quantum_1991}. Often mechanical models with a finite number of degrees of freedom are studied in relational scenarios to exhibit mathematical and conceptual subtleties within a fully relational quantum theory~\cite{ dewitt_quantum_1967, rovelli_quantum_1991, rovelliQuantumGravity2004,ashtekar_background_2004,Banerjee:2011qu,Bojowald:2010qpa,Gielen:2021igw,poulinToyModelRelational2006, bartlettReferenceFramesSuperselection2007,angeloPhysicsQuantumReference2011}. However, given that general relativity is a field theory, a natural stepping stone is a relational theory of quantum fields. 

With this as an aim, Dirac~\cite{diracLecturesQuantumMechanics1964} and later Kucha\v{r} \cite{Kuchar1973} (see also \cite{Isham:1984rz,Isham:1984sb,kuchar_parametrized_1989,*kuchar_dirac_1989}) developed what is known as parameterized field theory (PFT), in which a matter field $\phi$ is described relative to arbitrary curvilinear coordinates $X^\mu(t,x)$ associated with a foliation of Minkowski space. For constant $t$, these coordinate functions are treated as dynamical fields that describe hypersurfaces $\Sigma$ as embeddings into Minkowski space, $\mathsf{X}: \Sigma \to \mathcal{M}$. This leads to an extended phase space description of the matter field $\phi$ and the spacetime embedding fields characterizing $\mathsf{X}$. The resulting phase space can then be quantized via the Dirac procedure, leading to a  quantum theory that treats $\phi$ and~$\mathsf{X}$ on equal footing~\cite{ashtekar_manifestly_1994,torre_quantum_1998,torre_functional_1999,ashtekar_polymer_2002,ashtekar_background_2004,varadarajan_dirac_2007,kieferQuantumGravity2012,anastopoulos_gravitational_2021}. This approach to field theory is distinct from standard treatments (e.g.,~Ref.~\cite{weinberg_quantum_1995}) in which position is treated as a classical background parameter alongside the time variable. In contrast, PFT promotes both the time variable and the position operator to embedding fields, which are then quantized.

Prior to the development of PFT, Tomonaga~\cite{tomonaga_relativistically_1946,koba_relativistically_1947} and Schwinger~\cite{schwinger_quantum_1948,*schwinger_quantum_1949} put forward an alternative approach to a manifestly covariant formulation of quantum field theory. Inspired by the desire for a Lorentz invariant framework and the need to address divergences  in quantum electrodynamics, they promoted the time-dependent quantum mechanical wave function to a wave functional $\ket{\psi_\phi[\mathsf{X}]}$ by replacing the time variable with a spacelike hypersurface $\mathsf{X}$ in Minkowski space. The dynamics of the theory is governed by a functional differential equation known as the Tomonaga-Schwinger equation, which characterizes changes of $\ket{\psi_\phi[\mathsf{X}]}$ under deformations of $\mathsf{X}$, analogous to how the Schr\"{o}dinger equation governs time evolution.

In this article, we develop a novel approach to a relational formulation of quantum field theory along the lines envisaged by Page and Wootters~\cite{pageEvolutionEvolutionDynamics1983,woottersTimeReplacedQuantum1984}. We begin with the Dirac quantization of PFT, which requires constructing Hilbert spaces characterizing both a matter field and the embedding fields. We then introduce sets of coherent states $\{\ket{\mathsf{X}}\}$ relative to  one-dimensional subgroups of spacetime diffeomorphisms. These coherent states are eigenstates of the quantized embedding field operators with eigenfunctions corresponding to coordinate functions defining definite hypersurfaces. The group structure of these coherent states is crucial for implementing the Page-Wootters formalism~\cite{woottersTimeReplacedQuantum1984,smithQuantizingTimeInteracting2017,smith_quantum_2020,hohn_trinity_2021, hohn_equivalence_2021, hamette_quantum_2020,de_la_hamette_perspective-neutral_2021}. Physical states $\ket{\Psi_{\rm phys}}$ characterizing the scalar field and embedding fields are constructed by promoting the classical phase space constraints to operators that annihilate $\ket{\Psi_{\rm phys}}$. Conditioning physical states on a specific embedding state $\ket{\mathsf{X}}$ defines a state of the scalar field $\ket{\psi_\phi[\mathsf{X}]} \ce \bra{\mathsf{X}} \otimes \hat{I}_\phi \ket{\Psi_{\rm phys}}$, which is to be interpreted as the state of $\phi$ relative to $\mathsf{X}$. The conditional state is a wave functional that is shown to satisfy the Tomonaga-Schwinger equation, thus demonstrating the equivalence of the Page-Wootters formalism to standard formulations of wave functional dynamics in quantum field theory. In particular, we show that the foliation-independence of the integrated Tomonaga-Schwinger equation, which is usually attributed to the microcausality condition satisfied by the stress-energy tensor of the scalar field, is seen to be independently derived from the gauge invariance of physical states (see Fig.~\ref{fig:IntegrationOfTS}).

Moreover, the embedding  states are used to build relational Dirac observables and a relational Heisenberg picture equivalent to a quantum deparametrization of the physical Hilbert space~\cite{hoehnHowSwitchRelational2018,vanrietveldeSwitchingQuantumReference2018,Hoehn:2019,vanrietveldeChangePerspectiveSwitching2020,hohn_trinity_2021,hohn_equivalence_2021,de_la_hamette_perspective-neutral_2021}, which give unitarily equivalent formulations to the Page-Wootters formalism analogous to the non-interacting quantum mechanical scenario~\cite{hohn_trinity_2021,hohn_equivalence_2021}. Finally, we present a field-theoretic analog of the quantum reference frame changes introduced by Giacomini et al.~\cite{giacominiQuantumMechanicsCovariance2019}, which allows us to describe a quantum field theory relative to a superposition of classical embeddings. We show that when transforming the conditional state of the scalar field to one defined relative to an embedding in a state that has support on hypersurfaces related by Bogoliubov transformations that are non-trivial combinations of creation and annihilation operators, changing embedding fields that characterize a reference frame can lead to a particle creation effect. This effect is a field-theoretic analog of the conversion between spatial superposition and entanglement under a change of quantum reference frame~\cite{giacominiQuantumMechanicsCovariance2019}.

Note that the embedding fields constitute a field-theoretic quantum reference frame; specifically, the embedding fields are quantum reference frames associated with the infinite-dimensional diffeomorphism group, just like the quantum clocks in the Page-Wootters formalism are temporal quantum reference frames associated with the one-dimensional group of time reparametrizations. Our work thus constitutes a field-theoretic extension of recent results about quantum reference frames, which rather focused on finite-dimensional groups and thereby a mechanical setting or symmetry reduced cosmological models \cite{smithQuantizingTimeInteracting2017,hoehnHowSwitchRelational2018,vanrietveldeSwitchingQuantumReference2018,Hoehn:2019,loveridgeSymmetryReferenceFrames2018,giacominiQuantumMechanicsCovariance2019,vanrietveldeChangePerspectiveSwitching2020, hamette_quantum_2020,ruizQuantumClocksTemporal2020, smith_quantum_2020, hohn_trinity_2021,hohn_equivalence_2021,giacomini_spacetime_2021,de_la_hamette_perspective-neutral_2021,delaHamette:2021piz, ballesteros_group_2021,ali_ahmad_quantum_2022,carette_operational_2023,glowacki_quantum_2023,barbado_unruh_2020,Suleymanov:2023wio} (see also \cite{giacomini_second-quantized_2022,Kabel2022spacetimesuperpos,Goeller:2022rsx,Carrozza:2022xut,Kabel:2023jve} for dynamical frames in field-theory). 

While a functional Schr\"odinger equation governing the dynamics of a scalar field relative to variations of the embedding fields has previously been derived in the context of the canonical quantization of PFT~\cite{*kuchar_dirac_1989,torre_quantum_1998,torre_functional_1999,varadarajan_dirac_2007,kaya_schrodinger_2022}, this was done directly using the constraints on the physical Hilbert space. The novelty of our approach is that we derive a functional evolution equation on a reduced Hilbert space characterizing the scalar field alone in the form of the well-known Tomonaga-Schwinger equation from a quantum reference frame perspective and specifically the Page-Wootters formalism extended to field theory.

We begin in Sec.~\ref{sec: PFT} by reviewing parameterized field theory and derive a family of constraints satisfied by the scalar field and a set of embedding fields. In Sec.~\ref{sec: DiracPFT}, we construct the kinematical Hilbert spaces of the scalar field and embedding fields (the latter at a somewhat formal level) and carry out the Dirac quantization of PFT. In doing so, we introduce group coherent states associated with diffeomorphisms of Minkowski space. In Sec.~\ref{sec: relativematter}, we develop the field-theoretic generalization of the Page-Wootters formalism and show that the conditional state of the scalar field satisfies the Tomonaga-Schwinger equation, and construct relational Dirac observables encoding the same dynamics. We also develop a quantum deparametrization of the physical Hilbert space, leading to a relational Heisenberg picture. As an application of the developed framework, in Sec.~\ref{Sec: QETransformation} we analyze changes of quantum reference frames, induced Bogoliubov transformations on a scalar field, and a frame-dependent particle creation effect. We conclude in Sec.~\ref{sec: discussion} with a summary of our results and an outlook on future work.

Throughout, we adopt units such that $\hbar = c = 1$, employing the $(-, +)$ convention for the metric signature. We use round brackets ( ) for the arguments of functions and square brackets [ ] for the arguments of functionals.

\section{Parametrized field theory}
\label{sec: PFT}

Consider a real scalar field $\phi(X^\mu)$ living on 1+1 dimensional Minkowski space $\mathcal{M}$ equipped with the metric $\eta_{\mu\nu} = \diag(-1,1)$, defined in terms of the inertial coordinates $X^\mu =(T,X)$. The dynamics of the theory is specified by the action
\begin{equation}
S[\phi] = \int dT dX \, \mathcal{L}(\phi, \partial \phi / \partial X^\mu ),
\label{action1}
\end{equation}
where $\mathcal{L}(\phi, \partial \phi / \partial X^\mu )$ is the Lagrangian density of the field. One can vary this action and obtain the dynamics of the field with respect to the inertial coordinates~$X^\mu$. 

\begin{figure}[t]
\includegraphics[width= 0.45\textwidth]{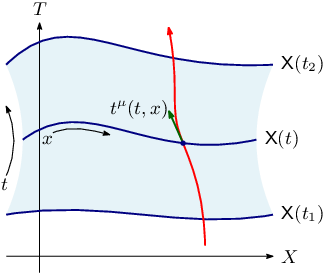}
\caption{Depicted is a smooth one-parameter family of spacelike hypersurfaces $\mathsf{X}(t)$ that define a foliation of Minkowski space, $\mathcal{F} \ce \{ \mathsf{X}(t), \ \forall t \in [t_1,t_2] \}$. The parameter $x$ is a spatial coordinate parameterizing each hypersurface (blue), and $t^\mu(t,x)$ denotes the deformation vector field (green), defined in Eq.~\eqref{eq: foliation}, which is tangent to integral curves (red) associated with the~$\mathcal{F}$.}
\label{foliationFig}
\end{figure}

However, the dynamics of the theory may be described with respect to an arbitrary set of curvilinear coordinates $x^\mu = (t,x)$ defined by the coordinate functions $X^\mu(t,x)$. Such a coordinate system can be chosen so that for each $t \in [t_1,t_2]$, the functions $X^\mu(t,x)$ parameterize a family of spacelike embeddings of the 1-manifold $\Sigma\simeq \mathbb{R}$ in $\mathcal{M}$:
\begin{equation*}
\begin{split}
\label{foliation1}
    \mathsf{X}(t) : \Sigma &\to \mathcal{M}, \\
    x &\mapsto X^\mu(t,x).     
\end{split}
\end{equation*}
 A foliation of Minkowski space is defined to be a smooth, one-parameter family $\mathcal{F} \ce \{ \mathsf{X}(t), \ \forall t \in [t_1,t_2] \}$ of spacelike hypersurfaces $\mathsf{X}(t)$ such that each spacetime point is located on precisely one hypersurface of the family; see Fig.~\ref{foliationFig}. A spacelike foliation defines a timelike deformation vector field~\cite{Isham:1984sb,thiemannModernCanonicalQuantum2008,wald2010general}
\begin{align}
    \label{eq: foliation}
    t^\mu(t,x) \ce \partial_t X^\mu(t,x) = N(t,x)n^\mu(t,x)+N^\mu(t,x),
\end{align}
where we have decomposed this vector field into its components orthogonal and parallel to the embedding $\mathsf{X}(t)$. Here, $n^\mu(t,x)$ is the unique future-pointing unit normal vector to $\mathsf{X}(t)$, $N(t,x)$ is known as the lapse function,  and $N^\mu(t,x)\ce N_x(t,x)\partial_x X^\mu(t,x)$ is the shift vector, which is orthogonal to $n^\mu(t,x)$. The flow generated by the deformation vector field defines a timelike congruence of curves in $\mathcal{M}$, which can be interpreted as a family of world lines corresponding to a set of observers relative to which the dynamics of the field is to be described. Also note that the timelike deformation vector field  is a complete vector field.\footnote{To see this, consider the integral curve $\gamma(t):(a,b)\rightarrow\mathcal{M}$ given by integrating the timelike deformation vector. Then, $\forall p\in\mathcal{M}$ and any integral curve $\gamma$ passing through the point $p$, $\gamma$ can be extended to an integral curve defined on $\mathbb{R}$, $\Gamma:\mathbb{R}\rightarrow\mathcal{M}$, and $\Gamma$ has a global flow.}

The action $S[\phi]$ can be re-cast on an extended configuration space characterizing both the field $\phi(t,x)$ and the embedding fields ${X}^\mu(t,x)$ as dynamical degrees of freedom \cite{Kuchar1973,kuchar_parametrized_1989,Isham:1984rz,Isham:1984sb}. This is accomplished by reparameterizing the action in terms of the coordinates $x^\mu$ as 
\begin{align}
S[\phi,X^\mu] = \int dt dx \, J \mathcal{L}\left(\phi, \frac{\partial \phi}{  \partial x^\nu} \frac{\partial x^\nu}{  \partial X^\mu}  \right),
\label{action2}
\end{align}
where $J \ce \frac{\partial T}{\partial t}\frac{\partial X}{\partial x}-\frac{\partial X}{\partial t}\frac{\partial T}{\partial x}$ denotes the determinant of the Jacobian associated with the coordinate transformation  $X^\mu(t,x)$. When cast in Hamiltonian form, the Lagrangian density in Eq.~\eqref{action2} yields the family of constraints~\cite{Kuchar1973,kuchar_parametrized_1989,Isham:1984rz,Isham:1984sb,Kuchar1982_conditional_PFT,kieferQuantumGravity2012}
\begin{align}
C_\mu(t,x) \ce \Pi_{\mu}(t,x) + h_\mu \left[ \phi, \pi, X^\mu \right](t,x) \approx 0,
\label{constraint}
\end{align}
where $\Pi_{\mu}(t,x)$ is the momentum conjugate to the inertial coordinate $X^\mu(t,x)$ and 
\begin{equation}
\begin{split}
    h_\mu \left[ \phi, \pi, X^\mu \right](t,x) &\ce J \frac{\partial t(t,x)}{\partial X^\nu} {T^\nu}_\mu(t,x)   \\
    &=  n_\nu(t,x) {T^\nu}_\mu(t,x), \label{HamFlux}
\end{split}
\end{equation}
 where $n_\nu(t,x)\ce  \epsilon_{\nu\rho} \partial X^\rho(t,x)/\partial x$ and  ${T^\mu}_\nu(t,x)$ denotes the stress-energy tensor of the field $\phi$.

\section{Dirac Quantization of PFT}
\label{sec: DiracPFT}

While it is true that the gauge-invariant dynamics defined by the action in Eq.~\eqref{action2} coincides with the dynamics defined by the action Eq.~\eqref{action1}, the former is defined on an extended configuration space that explicitly includes degrees of freedom associated with an embedding relative to which the scalar field is to be described. However, the embedding degrees of freedom are not independent of the matter field degrees of freedom due to the constraints in Eq.~\eqref{constraint}. 

Quantization proceeds via the Dirac prescription by promoting the configuration variables, $X^\mu(x)$ and $\phi(x)$, and their conjugate momenta, $\Pi_\mu(x)$ and $\pi(x)$, to operators satisfying canonical commutation relations on the kinematical Hilbert space $\mathcal{H}_{\rm kin} \simeq \mathcal{H}_{\rm Emb} \otimes \mathcal{H}_\phi$, where $\mathcal{H}_{\rm Emb}$ and $\mathcal{H}_\phi$ are the kinematical embedding and matter field Hilbert spaces, respectively. These canonical phase space  operators then define a quantization of the constraint operator in Eq.~\eqref{constraint} upon a choice of factor ordering and regularization. Physical states are defined as those that are annihilated by the constraint operator and used to construct the physical Hilbert space $\mathcal{H}_{\rm phys}$. We outline the formal construction of $\mathcal{H}_\phi$, $\mathcal{H}_{\rm Emb}$, and $\mathcal{H}_{\rm phys}$, and introduce structures necessary for a field-theoretic generalization of the Page-Wootters formalism.

\subsection{The matter field Hilbert space $\mathcal{H}_\phi$}
\label{fieldHilbertSpace}

As a minimal assumption, take the classical configuration space of the matter field $\mathcal{C}_\phi = \{\phi(x)\}$ to be the space of twice differentiable functions that decay rapidly at infinity on the hypersurface $\mathsf{X}$. By analogy with the quantum description of a nonrelativistic free particle, where the Hilbert space is the space of square-integrable functions over the configuration space $\mathbb{R}^3$, one would na\"ively expect that the Hilbert space of the matter field to be $\mathcal{H}_\phi \simeq L^2(\mathcal{C}_\phi, \mathcal{D}\phi)$ for some measure $\mathcal{D}\phi$. However, the field configuration space $\mathcal{C}_\phi$ is infinite-dimensional, complicating the integration theory necessary to define  $\mathcal{H}_\phi$. To address these issues, we follow the construction of $\mathcal{H}_{\phi}$ presented in Refs.~\cite{glimm_quantum_1981,ashtekar_manifestly_1994,ashtekar_background_2004,ticciati_quantum_1999}.

One proceeds by introducing an arbitrary, linearly independent set of probe functions $V = \{v_1(x), \dots, v_n(x)\}$ that are elements of the Schwartz space $\mathcal{S}$ of smooth, rapidly decreasing functions on $\mathbb{R}$. These functions probe the structure of the field configuration space through the linear functionals constructed by smearing the field operators on $\Sigma$ 
\begin{equation*}
    \phi_i \ce \int d\Sigma(x) \,  v_i(x) \phi(x),
\end{equation*}
where $d\Sigma(x) \ce dx \sqrt{\gamma(x)}$ is the invariant measure on $\Sigma$ and $\gamma(x)\ce \eta_{\mu\nu}\partial_x X^\mu(x) \partial_x X^\nu(x)$ is both the induced metric on $\Sigma$ and its determinant; the numbers $\phi_i$ can be interpreted as the component of the field along $v_i$. One then defines the set $\mathrm{Cyl}_V$ of {so-called} cylindrical functions on $\mathcal{C}_\phi$ with respect to $V$ as those that can be expressed as functions of the coordinates $\phi_i$,
\begin{equation}
    f(\phi) = f(\phi_1, \dots, \phi_n) \in \mathrm{Cyl}_V,
    \label{cylV}
\end{equation}
i.e.\ functions that are constant along the other directions in $\mathcal{C}_\phi$ (not encompassed by the probes), and introduces an inner product on $\mathrm{Cyl}_V$
\begin{equation}
    \braket{f, g} \ce \int d\mu_n \, f^*(\phi_1, \dots, \phi_n) g(\phi_1, \dots, \phi_n),
    \label{innerProductCylV}
\end{equation}
where $d\mu_n$ is a measure on $\mathbb{R}^n$. The space $\mathrm{Cyl}_V$ is then extended to the space $\mathrm{Cyl}$ of all functions that are cylindrical with respect to some set of probes, not necessarily the set $V$. For the inner product in Eq.~\eqref{innerProductCylV} to be well-defined, the value of the integral cannot depend on the specific set of probes used to represent the functions. This imposes nontrivial consistency requirements on the measure $d\mu_n$ on $\mathrm{Cyl}$ that can be met, for example, by a normalized Gaussian measure on $\mathbb{R}^n$. The Cauchy completion of $\mathrm{Cyl}$ with respect to the norm induced by Eq.~\eqref{innerProductCylV} is taken to be the Hilbert space of the field $\mathcal{H}_\phi$. One can then extend the measure $\mathcal{D}\phi \ce \lim_{n \to \infty} d\mu_n$ onto the space of tempered distributions $\mathcal{S}'$, which is the topological dual of the Schwartz space of probes $\mathcal{S}$ (i.e., the space of linear functionals on $\mathcal{S}$), and show that $\mathcal{H}_\phi \simeq L^2(\mathcal{S}',\mathcal{D}\phi)$; for this reason, $\mathcal{S}' \supset \mathcal{C}_\phi$ is known as the quantum configuration space, which is larger than $\mathcal{C}_\phi$~\cite{ashtekar_manifestly_1994,ashtekar_background_2004}. In that way, it encodes the distributional character of quantum field theory.

Having defined the Hilbert space $\mathcal{H}_\phi$, we seek a representation of the field operator, its conjugate momentum, and their canonical commutation relations. Analogous to the generalized eigenstates of the position operator in nonrelativistic quantum mechanics, we can define generalized field eigenstates $\ket{\phi}$ associated with configurations of the field $\phi(x) \in \mathcal{S}'$. More specifically, consider the distributional field operator $\hat{\phi}(x)$ on a hypersurface $\mathsf{X}$ satisfying
\begin{equation*}
    \hat{\phi}(x) \ket{\phi} = \phi(x) \ket{\phi}.
\end{equation*}
The states $\ket{\phi}$ can be represented as a delta functional on~$\mathcal{S}'$,
\begin{equation*}
    \ket{\phi} = \int_{\mathcal{S}'} \mathcal{D}\phi' \, \delta[\phi - \phi'] \ket{\phi'},
\end{equation*}
and are orthogonal to one another $\braket{\phi|\phi'} = \delta [\phi - \phi']$. It follows that these states form a basis for $\mathcal{H}_\phi$, so that any field state on $\mathsf{X}$ may be expanded as
\begin{equation*}
    \ket{\Psi_\phi} = \int_{\mathcal{S}'}  \mathcal{D}\phi \, \Psi_\phi[\phi] \ket{\phi},
\end{equation*}
where $\Psi_\phi[\phi] = \braket{\phi| \Psi_\phi} \in \mathcal{H}_\phi $. 

It will be useful to smear the distributional field operator with a probe function $f \in \mathcal{S}$,
\begin{equation*}
    \hat{\phi}[f] \ce  \int d\Sigma(x) \,  f(x) \hat{\phi}(x),
\end{equation*}
 and define its conjugate momentum 
\begin{equation*}
    \hat{\pi}[g] 
    \ce -i\int_{\mathcal{S}'} \mathcal{D}\phi \int d\Sigma(x) \,  g(x) \ket{\phi} \frac{\delta}{\delta \phi(x)} \bra{\phi},
\end{equation*}
which together furnish a representation of the canonical commutation relations
\begin{equation*}
\left[ \hat{\phi}[f], \hat{\pi}[g]\right] \ket{\Psi_\phi} = i \left( \int  d\Sigma(x)   f(x)g(x) \right) \ket{\Psi_\phi}.
\end{equation*}

\subsection{The embedding field Hilbert space $\mathcal{H}_{\rm Emb}$}

\subsubsection{General framework}

The embedding field Hilbert space $\mathcal{H}_{\rm Emb}$ ought to be defined as the space of square-integrable functions over the configuration space  of the embedding fields $Q$ containing all possible hypersurfaces, $\mathsf{X}:\Sigma \to \mathcal{M} \in \text{Emb}(\Sigma, \mathcal{M})$. This is a mathematically subtle topic because $\text{Emb}(\Sigma,\mathcal{M})$ is not a vector space (one cannot add arbitrary embeddings to form another one), and we shall invoke a few assumptions to proceed in analogy to the construction of the scalar field Hilbert space.\footnote{For an alternative construction using polymer quantization, see~\cite{varadarajan_dirac_2007}.} Following Isham and Kucha\v{r}~\cite{Isham:1984sb,Isham:1984rz}, we first consider the space of all (not necessarily spacelike) embeddings $\text{Emb}: \Sigma \to \mathcal{M}$, where $\Sigma = \mathbb{R}$. We then take the configuration space to be 
\begin{equation*}
    Q \ce \text{Emb}(\Sigma, \mathcal{M})/\text{Diff} (\Sigma) \simeq  G/H,
\end{equation*}
where $G \ce \text{Diff} ( \mathcal{M})$ is the group of diffeomorphisms on $\mathcal{M}$, $H \ce \text{Diff}(\mathcal{M},\mathsf{X}_0)$ is the subgroup of diffeomorphisms that leave invariant a fiducial embedding $\mathsf{X}_0: \Sigma \to \mathcal{M}$, and $\text{Diff}(\Sigma)$ is the group of diffeomorphisms on $\Sigma$. The embedding field Hilbert space will then be $\mathcal{H}_{\rm emb} \simeq L^2(\tilde{Q},\mathcal{D}q)$  for an appropriate measure $\mathcal{D}q$ on the quantum configuration space $\tilde{Q} \supset Q$.\footnote{In arriving at this point, we assume that it is possible to define a complete orthonormal set of probe one-forms $B_q\ce\{q_{1\mu}(x),q_{2\mu}(x),\dots,q_{n\mu}(x)\}$ as elements of the Schwartz space $\mathcal{S}(\Sigma)$ of rapidly decreasing functions on $\Sigma$. Proceeding analogously to Sec.~\ref{fieldHilbertSpace}, let us introduce  linear functions that probe the configuration space
\begin{equation}
    q_i :=\int d\Sigma(x) \,   q_{i\mu}(x)X^\mu(x) \in \mathbb{R}. \nn
\end{equation}
We assume that one can again define the set of cylindrical functions $\mathrm{Cyl}_{B_q}$ with respect to $B_q$, 
\begin{equation}
\psi(q) = \psi(q_1,q_2,\dots,q_n) \in \mathrm{Cyl}_{B_q},   \nn  
\end{equation}
analogous to Eq.~\eqref{cylV}, and introduce an inner product on $\mathrm{Cyl}_{B_q}$ analogous to Eq.~\eqref{innerProductCylV}, which requires the introduction of a cylindrical measure $d\mu_n =\prod_{i=1}^n d\mu$, which we take to be constructed as a product measure. We assume that this space can be extended to the space of all functions on $Q$ that are cylindrical with respect to some set of probe functions, not necessarily $B_q$, and then take the Cauchy completion of this space to form the embedding Hilbert space~$\mathcal{H}_{\rm Emb}$. We further assume that the embedding Hilbert space can be represented as $\mathcal{H}_{\rm Emb} \simeq L^2(\tilde Q,\mathcal{D} q)$, where $\mathcal{D} q \ce \lim_{n \to \infty} d \mu_n$ and $\tilde{Q} = \mathcal{S}_2'(\Sigma)$ is a suitable space of two-component tempered distributions on $\Sigma$.}

 The embedding fields are assumed to be promoted to self-adjoint operators on $\mathcal{H}_{\rm Emb}$ defined by the eigenvalue-eigenvector equation 
\begin{equation*}
    \hat{X}^\mu(x) \ket{\mathsf{X}_{q}} \ce X_q^\mu(x) \ket{\mathsf{X}_{q}}, 
\end{equation*}
where $X_q^\mu(x): \Sigma \to \mathcal{M}$ is the eigenfunction corresponding to the generalized eigenstate $\ket{\mathsf{X}_{q}}$ and equal to the coordinate functions describing the embedding $\mathsf{X}_q$.

Given that the embedding field operators are taken to be self-adjoint,  the embedding operator $\hat{X}^\mu(x)$ is densely defined on the basis $\ket{\mathsf{X}_q}$, which are generalized eigenstates of the embedding operator $\hat{X}^\mu(x)$ with real eigenvalue functions $X^\mu_q(x)$. This is consistent with the demand that $\hat{X}^\mu(x)$ is self-adjoint and thus $\ket{\mathsf{X}_{q}}$ are orthogonal for different $q$, so that we may represent hypersurface states as Dirac delta functionals
\begin{equation*}
\ket{\mathsf{X}_q} = \int \mathcal{D} q'\, \delta [\mathsf{X}_{q'} - \mathsf{X}_q ] \ket{\mathsf{X}_{q'}}.
\end{equation*}
It follows that $\{\ket{\mathsf{X}_{q}}, \ \forall q \in \tilde{Q}\}$ forms a basis for $\mathcal{H}_{\rm Emb}$, and thus we have a resolution of the identity
\begin{equation}\label{resolid}
    \hat{I}_{\rm Emb} = \int \mathcal{D}q \, \ket{\mathsf{X}_{q}}\!\bra{\mathsf{X}_{q}},
\end{equation}
and we can expand embedding states as
\begin{equation*}
   \ket{\psi} =  \int \mathcal{D}q \, \psi[\mathsf{X}_q] \ket{\mathsf{X}_{q}},
\end{equation*}
where $\psi[\mathsf{X}_q]  \ce \braket{\mathsf{X}_{q}|\psi}$.

The embedding operators $\hat{X}^\mu(x)$ may be smeared with a one-form $w_{\mu}(x)$ on $\Sigma$, yielding 
\begin{align*}
\hat{\mathsf{X}}[w] &\ce \int d\Sigma(x) \,w_\mu(x) \hat{X}^\mu(x) \nn \\
&= \int d\Sigma(x) \, w_\mu(x) \int \mathcal{D} q \, \ket{\mathsf{X}_q} {X}_q^\mu(x)\bra{\mathsf{X}_q}.
\end{align*}

The momentum operator canonically conjugate to $\hat{\mathsf{X}}[w]$ can be represented as
\begin{align*}
    \hat{\mathsf{\Pi}}[v] & \ce  \int d\Sigma(x) \, v^\mu(x) \hat{\Pi}_\mu(x) \nn \\
    &= -i\int d\Sigma(x) \, v^\mu(x) \int \mathcal{D} q   \ket{\mathsf{X}_q} \frac{\delta}{\delta X_q^\mu(x)} \bra{\mathsf{X}_q},
\end{align*}
where $v^\mu(x)$ is a vector field on $\Sigma$. Together, these operators furnish a representation of the canonical commutation relations
\begin{equation*}
    \left[ \hat{\mathsf{X}}[w], \hat{\mathsf{\Pi}}[v]\right] 
     = i  \int d \Sigma(x)\, w_\mu(x) v^\mu(x) .
    \label{EmbeddingCCR}
\end{equation*}
Note that $\hat{\mathsf{\Pi}}[v]$ cannot be a self-adjoint operator in general. If this were the case, then the canonical commutation relation would imply that the embedding configuration space has a linear structure and would thus be a vector space. However, this is not true, as one cannot add arbitrary embeddings to produce another embedding (see \cite{Isham:1984rz,Isham:1984sb} for further discussion). Nevertheless, because $\hat{\mathsf{\Pi}}[v]$ is symmetric, for different vector fields $v^\mu$ it will have non-trivial domains in $\mathcal{H}_{\rm emb}$, where it may effectively act like a self-adjoint operator. On these domains it will act like a translation operator for the embedding fields, however, only locally in the embedding configuration space.

\subsubsection{Coherent states relative to subgroups of spacetime diffeomorphisms}

\begin{figure}
    \centering
    \includegraphics[width=0.45\textwidth]{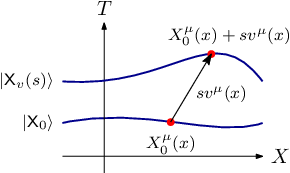}
    \caption{Hypersurfaces corresponding the  seed state $\ket{\mathsf{X}_0}$ and the state $\ket{\mathsf{X}_{v}(s)} \ce \hat{U}_v(s)\ket{\mathsf{X}_0}$, which is connected to the seed embedding by a unitary representation $\hat{U}_v(s)$ of the one-dimensional subgroup $G_v \subset \mathrm{Diff}(\mathcal{M})$ generated by the vector field $v^\mu(x) \in \mathrm{Vec}(\mathcal{M})$.}
    \label{fig:EmbeddingStates}
\end{figure}

Consider a representation of a one-dimensional subgroup  $G_v \subset  \mathrm{Diff}(\mathcal{M})$ generated by the vector field $v^\mu(x) \in \mathrm{Vec}(\mathcal{M})$, which is taken to vanish asymptotically, given by 
\begin{equation}
    \hat{U}_v(s) \ce  e^{-is \hat{\mathsf{\Pi}}[v]} = e^{-is \int d\Sigma(x) \, v^{\mu}(x) \hat{\Pi}_\mu(x)},
    \label{unitaryRep}
\end{equation}
for all $s \in [s_0,s_1]\subset\mathbb{R}$, where this interval will depend on the states this operator acts on. For some (possibly distributional) states it may even be defined for all $s\in\mathbb{R}$. On the domain of $\hat{\mathsf{\Pi}}[v]$, this group representation will effectively act unitarily. Let us now consider a seed state $\ket{\mathsf{X}_0}$, taken to be a generalized eigenstate of the embedding field operator which thus corresponds to a definite hypersurface configuration, and define the set of group coherent states~\cite{perelomovGeneralizedCoherentStates1986} generated by the action of $\hat{U}_v(s)$:
\begin{equation}
    \left\{\ket{\mathsf{X}_{v}(s)} \ce \hat{U}_v(s )\ket{\mathsf{X}_0} \mbox{ for all } s\in[s_0,s_1]\subset\mathbb{R} \right\}\,,
    \label{DefCoherntStates}
\end{equation} 
where the interval $[s_0,s_1]$ depends on $\mathsf{X}_0$.
Similarly defined coherent states relative to subgroups of the diffeomorphism group of Euclidean space have found application in general quantization methods~\cite{twareque_ali_quantization_1991}. As shown in Appendix~\ref{ProofOfTheorems}, these coherent states satisfy
\begin{align}
     \hat{X}^{\mu}(x) \ket{\mathsf{X}_{v}(s)}  &= \Big(   X_0^\mu(x) + sv^\mu(x) \Big) \ket{\mathsf{X}_{v}(s)}.
    \label{coherntEigenTry2}
\end{align}
This yields an interpretation of $\ket{\mathsf{X}_{v}(s)}$ as a state of the embedding fields associated with a hypersurface described by the coordinate functions $X^\mu(x) = X_0^\mu(x) + sv^\mu(x)$, where $X_0^\mu(x)$ are the coordinate functions of the seed hypersurface $\mathsf{X}_0$; see Fig.~\ref{fig:EmbeddingStates}.

\subsection{The physical Hilbert space $\mathcal{H}_{\rm phys}$}

Physical states $\ket{\Psi_{\rm phys}}$ are defined as those living in the kernel of the constraint operators resulting from the quantization of Eq.~\eqref{constraint}:\footnote{Note that $\hat{C}_\mu$ can in general not be a self-adjoint operator on $\mathcal{H}_{\rm kin}$ due to $\hat\Pi_\mu$ not being self-adjoint. However, one can still impose it to construct its space of solutions.}
\begin{equation}
     \hat{C}_\mu(x) \ket{\Psi_{\rm phys}} = \left( \hat{\Pi}_\mu(x) + \hat{h}_\mu(x)   \right) \ket{\Psi_{\rm phys}} =  0,
\label{QuantumConstraint}
\end{equation}
where
\begin{equation}
    \hat{h}_\mu(x) \ce  \hat{n}_\nu(x)\otimes  {\hat{T}^{\nu}}_{\ \mu}(x)  + \hat{A}_\mu(x) 
    \label{systemHamiltonianPhi}
\end{equation}
 is the quantization of the Hamiltonian flux in Eq.~\eqref{HamFlux}, which picks up an anomalous term 
\begin{equation*}
    \hat{A}_\mu(x) \!=\! -\! \int \mathcal{D}q \,  \ket{\mathsf{X}_q} \sqrt{\gamma(x)}\! \left( \!\eta_{\mu \nu} X_q^\nu(x) \frac{\partial\mathcal{K}}{\partial x}\!-\!\mathcal{K}^2n_\mu \!\right) \!\bra{\mathsf{X}_q},
\end{equation*}
if we consider a free scalar field theory, i.e.\ one with mass $m=0$. In that case, we are dealing with a conformal field theory (CFT) which leads to anomalies in the quantum setting. Adding the anomaly to the Hamiltonian ensures the constraint algebra closes~\cite{kuchar_dirac_1989}, and is further discussed in Appendix~\ref{anomaly}. In the expression above, $\mathcal{K}$ represents the trace of the extrinsic curvature of the embedding $\mathsf{X}_q$. The effect of this anomalous term is to redistribute the energy content of the slice based on how it is embedded in the higher dimensional spacetime \cite{kuchar_dirac_1989}. Notice that when acting on states with support only on flat embeddings, the anomalous term vanishes, as such embeddings have constant zero extrinsic curvature.  One should also keep in mind that if the embeddings are taken to be flat at infinity, the anomaly vanishes when integrated over the entire spacelike slice (in any coordinate frame)\footnote{In \cite{kuchar_dirac_1989}, this was shown for cylindrical Minkowski space, but from the equations in that work it follows that this also holds for open Minkowski on hypersurfaces that become flat towards infinity.} 
\begin{equation}
\label{eq: vanishanomaly}
    \int d\Sigma_q(x)\, \hat{A}_\mu(x)=0.
\end{equation}
If the embedding states $\ket{\mathsf{X}_v(s)}$ in Eq.~\eqref{DefCoherntStates} are generated from an asymptotically flat embedding $\mathsf{X}_0$, then the anomaly vanishes when integrated over the associated hypersurface $\mathsf{X}_v(s)$.

We assume that the set of states satisfying Eq.~\eqref{QuantumConstraint} can be equipped with an inner product and completed to form the physical Hilbert space~$\mathcal{H}_{\rm phys}$. Gauge transformations are generated by $\hat{\mathsf{C}}[v] \ce \hat{\mathsf{\Pi}}[v] + \hat{\mathsf{h}}[v]$ and leave physical states invariant, $e^{i\hat{\mathsf{C}}[v]}\ket{\Psi_{\rm phys}} = \ket{\Psi_{\rm phys}}$, for all $\ket{\Psi_{\rm phys}} \in \mathcal{H}_{\rm phys}$.

\section{Matter relative to quantum fields}
\label{sec: relativematter}
When reparametrization-invariant theories are canonically quantized, it is found that the Hamiltonian of the theory is proportional to a constraint operator. This implies that physical states do not evolve under the action generated by the quantized Hamiltonian.  This is a generic feature of relativistic theories and, ultimately, a manifestation of background independence. This means that there is no physical evolution relative to the external background.

Nonetheless, we are tasked with recovering a notion of dynamics from the quantized theory. One approach is to identify a subsystem after quantization to serve as a reference frame relative to which another subsystem evolves. Three approaches that accomplish this task are the construction of so-called relational Dirac observables (a.k.a., evolving constants of motion)~\cite{rovelliQuantumGravity2004,dittrichPartialCompleteObservables2006a,dittrichPartialCompleteObservables2007a,thiemannModernCanonicalQuantum2008}, the Page and Wootters formalism~\cite{pageEvolutionEvolutionDynamics1983,woottersTimeReplacedQuantum1984}, and a quantum deparametrization procedure~\cite{hoehnHowSwitchRelational2018,Hoehn:2019}. For simple mechanical systems, these approaches have been shown to be equivalent in \cite{hohn_trinity_2021,hohn_equivalence_2021}.

In this section, we introduce the Page-Wootters formulation of the parameterized field theory reviewed in Sec.~\ref{sec: PFT}. A conditional state of the matter field is defined by conditioning physical states on configurations of the embedding fields. This conditional state is then shown to formally satisfy the covariant Tomonaga-Schwinger equation and an appropriate Schr\"{o}dinger equation describing evolution along arbitrary spacelike foliations of Minkowski space, thus demonstrating the formal equivalence of the Page-Wootters formalism with standard formulations of quantum field theory. We also introduce relational Dirac observables and a quantum deparameterization of the physical Hilbert space that results in a relational Heisenberg picture. These relational formulations are shown to be unitarily equivalent to the Page-Wootters formalism.

\subsection{The Page-Wootters formalism and Tomonaga-Schwinger equation}
\label{PWconstruction}

The typical starting point of the Page-Wootters formalism applied to a mechanical system is a physical state $\ket{\Psi_{\rm phys}}$ describing a non-interacting clock $C$ and system of interest $S$ satisfying a Hamiltonian constraint 
\begin{equation}
    \hat{C}_H \ket{\Psi_{\rm phys}} = \left(\hat{H}_C+ \hat{H}_S \right)  \ket{\Psi_{\rm phys}} = 0.
    \label{PWconstraint1}
\end{equation}
One then considers a time observable 
\begin{equation*}
   T_C \ce \{ \ket{t}\!\bra{t}, \ \forall t \in G \subseteq \mathbb{R} \},
\end{equation*}
defined by a set of rank-1 effect density operators $\ket{t}\!\bra{t}$ constructed from the outer product of so-called clock states $\ket{t}$ for all $t \in \mathbb{R}$ that parameterize the one-dimensional time evolution group $G$ generated by the clock Hamiltonian $\hat{H}_C$. What distinguishes $T_C$ as a time observable is that it transforms covariantly under the action of $G$, which amounts to the following relation between the clock states\footnote{We note that in the case when the spectrum of $\hat{H}_C$ is unbounded, then $T_C$ can be associated with a self-adjoint time operator $\hat{T}_C$ that is canonically conjugate to $\hat{H}_C$, ensuring that the clock Hamiltonian generates time translations. However, in the case when $\hat{H}_C$ is  bounded from below, as in any physical system, then such a construction is not possible. Nonetheless, the covariant POVM $T_C$ is still well-defined and yields the optimal estimate of parameter time~\cite{holevoProbabilisticStatisticalAspects1982,buschQuantumTheoryMeasurement1991,*buschOperationalQuantumPhysics1995,*buschQuantumMeasurement2016,smithQuantizingTimeInteracting2017,smith_quantum_2020,hohn_trinity_2021,hohn_equivalence_2021}.}
\begin{equation}
    \ket{t'} = e^{-i \hat{H}_C \left(t'-t\right)} \ket{t}.
    \label{PWcovariant}
\end{equation}
One then defines the conditional state of the system $\ket{\psi_S(t)}$ at a time $t$ as
\begin{equation}
    \ket{\psi_S(t)} \ce \left(\bra{t} \otimes \hat{I}_S \right) \ket{\Psi_{\rm phys}}.
    \label{PWconditionalState}
\end{equation}
It then follows from Eqs.~\eqref{PWconstraint1} and \eqref{PWcovariant} that the conditional wave function satisfies the Schr\"{o}dinger equation, $i\frac{d}{dt} \ket{\psi_S(t)} = \hat{H}_S \ket{\psi_S(t)}$, and yields the correct one- and two-time probabilities~\cite{hohn_trinity_2021}. Thus, we conclude that the conditional state $\ket{\psi_S(t)}$ ought to be interpreted as the usual time-dependent solution to the Schr\"{o}dinger equation, which constitutes the recovery of the standard quantum mechanical framework from the Page-Wootters formalism. 

However, Page and Wootters never intended these ideas to be limited to mechanical systems, which is made clear when Page writes~\cite{page_time_1989}:
\begin{quote}
    `For simplicity, one can think of the [\dots] system as a ``particle'' with ``position'', which is a convenient model to have in mind, but my discussion is not intended to be limited to such a simple model. For example, the closed system could consist of relativistic quantum fields within a given classical background spacetime.'
\end{quote}
Taking this seriously, we develop a field-theoretic generalization of the Page-Wootters formalism for PFT. We begin by considering a physical state $\ket{\Psi_{\rm phys}}$ satisfying Eq.~\eqref{QuantumConstraint}, and introducing the so-called reduction map $\mathcal{R}_{\mathbf S}[\mathsf{X}_q]$, which plays the same role as Eq.~\eqref{PWconditionalState}. The reduction map is defined as
\begin{equation}
\begin{split}
\label{eq: framereorient}
    \mathcal{R}_{\mathbf S}[\mathsf{X}_q]: \mathcal{H}_{\rm phys} &\rightarrow \mathcal{H}_{\mathsf{X}_q}, \nn \\ \ \ket{\Psi_{\rm phys}} &\mapsto \ket{\psi_\phi[\mathsf{X}_q]}   \ce \mathcal{R}_{\mathbf S}[\mathsf{X}_q] \ket{\Psi_{\rm phys}},
\end{split}
\end{equation}
where
\begin{equation}
\mathcal{R}_{\mathbf S}[\mathsf{X}_q] \ce \bra{\mathsf{X}_q}\otimes  \hat{I}_{\phi},
\label{defReduction}
\end{equation}
is used to condition a physical state $\ket{\Psi_{\rm phys}}$ on an embedding state $\ket{\mathsf{X}_q}$. We refer to $\ket{\psi_\phi[\mathsf{X}_q]}$ as a conditional state of the field $\phi$ relative to the hypersurface $\mathsf{X}_q$.

As a consequence of the resolution of the identity in Eq.~\eqref{resolid}, the reduction map is formally invertible on physical states, with inverse
\begin{align*}
    \mathcal{R}^{-1}_{\mathbf S}[\mathsf{X}_q]: \mathcal{H}_{\mathsf{X}_q} &\rightarrow \mathcal{H}_{\rm phys} ; \nn \\ \ \ket{\psi_\phi[\mathsf{X}_q]} &\mapsto \ket{\Psi_{\rm phys}}   = \mathcal{R}^{-1}_{\mathbf S}[\mathsf{X}_q] \ket{\psi_\phi[\mathsf{X}_q]},
\end{align*}
where
\begin{align}
    \mathcal{R}_{\mathbf S}^{-1}[\mathsf{X}_q] &= \int \mathcal{D} q' \, \ket{\mathsf{X}_{q'}} \otimes \hat{U}_{\phi}[\mathsf{X}_{q'} - \mathsf{X}_{q}] \label{defReductionInverse} \\
    &= \int \mathcal{D} q' \,  \hat{U}_{\mathsf{X}}[\mathsf{X}_{q'} - \mathsf{X}_{q}] \ket{\mathsf{X}_q} \otimes \hat{U}_{\phi}[\mathsf{X}_{q'} - \mathsf{X}_{q}] , 
    \nn
\end{align}
and 
\begin{align}
    \hat{U}_{\mathsf{X}}[\mathsf{X}_{q'} - \mathsf{X}_{q}] &\ce e^{-i \int d\Sigma_q(x) \, \left( X^{\mu}_{q'}(x) - X^{\mu}_{q}(x) \right) \hat{\Pi}_\mu(x) }, \nn \\
    \label{eq: fieldunitary}
    \hat{U}_{\phi}[\mathsf{X}_{q'} - \mathsf{X}_{q}] &\ce e^{-i \int d\Sigma_q(x) \, \left( X^{\mu}_{q'}(x) - X^{\mu}_{q}(x) \right) n_\nu(x) {\hat{T}^{\nu}}_{\ \mu}(x) }. 
\end{align}
Note that the surface integrals above are defined on the embedded hypersurface $\mathsf{X}_q$ in Minkowski space and that the difference $X^\mu_{q'}(x)-X^\mu_q(x)$ defines a vector field on (but not necessarily tangential to) this surface. In the line above, $n_\nu(x)$ is no longer an operator as it has already been evaluated on the embedding $\mathsf{X}_q$. The action of the inverse map on a conditional state of the field is then
\begin{equation}
    \mathcal{R}_{\mathbf S}^{-1}[\mathsf{X}_q] \ket{\psi_\phi[\mathsf{X}_q]} = \int \mathcal{D} q \,  \ket{\mathsf{X}_q}\ket{\psi_\phi[\mathsf{X}_q]} = \ket{\Psi_{\rm phys}}.\label{entangled}
\end{equation}
Given that $\mathcal{R}_{\mathbf S}[\mathsf{X}_q]$ is invertible, it follows that $\mathcal{H}_{\rm phys}$ is isomorphic to $\mathcal{H}_{\mathsf{X}_q}$.

We will now consider the conditional state of the field 
\begin{equation}
    \ket{\psi_\phi[\mathsf{X}_{v}(s)]} \ce  \mathcal{R}_{\mathbf S} [\mathsf{X}_{v}(s)] \ket{\Psi_{\rm phys}}.
\label{ConditionalWaveFunctional}
\end{equation}
associated with a one-dimensional subgroup $G_v \subset G$ defined by the vector field $v^\mu(x)$ and group element $s$. In what follows, we will derive the variation of $\ket{\psi_\phi[\mathsf{X}_{v}(s)]}$ in the parameter $s$, parameterizing the subgroup $G_v$, and then its variation in the vector field $v^\mu(x)$, characterizing deformations of the hypersurface $\mathsf{X}_0$. We then consider the evolution of the conditional state along a foliation $\mathcal{F}$ of $\mathcal{M}$.

\subsubsection{Evolution in the group parameter $s$}

One notable difference to the quantum mechanical case in Eq.~\eqref{PWconstraint1} is that in PFT the embedding fields, serving as a dynamical reference frame for the scalar field, interacts with the scalar field via the embedding-dependent Hamiltonian in Eq.~\eqref{systemHamiltonianPhi} (cf.,~Ref.~\cite{smithQuantizingTimeInteracting2017}). Nevertheless, we can proceed similarly. 

The evolution of the conditional state $\ket{\psi_\phi[\mathsf{X}_{v}(s)]}$ in the parameter~$s\in[s_0,s_1]$ satisfies:
\begin{equation}
\begin{split}
    i \frac{d}{ds} \ket{\psi_\phi[\mathsf{X}_{v}(s)]} &= i \frac{d}{ds}  \braket{\mathsf{X}_{v}(s) | \Psi_{\rm phys}}  \\
    &= i \frac{d}{ds} \bra{\mathsf{X}_0} e^{is\hat{\mathsf{\Pi}}[v]} \ket{\Psi_{\rm phys}}  \\
    &= i \bra{\mathsf{X}_0} e^{is\hat{\mathsf{\Pi}}[v]} i \hat{\mathsf{\Pi}}[v] \ket{\Psi_{\rm phys}}  \\
    &= - \bra{\mathsf{X}_{v}(s)} \hat{\mathsf{\Pi}}[v] \ket{\Psi_{\rm phys}}  \\
    &= \bra{\mathsf{X}_{v}(s)} \hat{\mathsf{h}}[v] \ket{\Psi_{\rm phys}}  \\
    &= \hat{\mathsf{h}}[v,\mathsf{X}_{v}(s)] \braket{\mathsf{X}_{v}(s) | \Psi_{\rm phys}}  \\
    &= \hat{\mathsf{h}}[v,\mathsf{X}_{v}(s)] \ket{\psi_\phi[\mathsf{X}_{v}(s)]},
\end{split}
    \label{parmeterEvolution}
\end{equation}
where the dependence on $\mathsf{X}_{v}(s)$ is through $n_\mu(x)$, and we have defined  
\begin{equation}
    \hat{\mathsf{h}}[v,\mathsf{X}_{v}(s) ] \ce \int d\Sigma(x) \, v^\mu(x)  n_\nu(x)  {\hat{T}^{\nu}}_{\ \mu}(x),
    \label{generator}
\end{equation}
and made use of Eqs.~\eqref{unitaryRep}, \eqref{QuantumConstraint}, \eqref{ConditionalWaveFunctional}. It is thus seen that the operator $\hat{\mathsf{h}}[v,\mathsf{X}_{v}(s) ]$ generates the flow in $s$. We remind the reader that the anomaly (when present) has vanished because we have integrated over asymptotically flat spatial hypersurfaces as in Eq.~\eqref{eq: vanishanomaly}.

As an example, let us consider an inertial foliation of Minkowski space by flat spacelike hypersurfaces defined by the one-parameter family of hypersurfaces through the coordinate functions
\begin{equation*}
    X^\mu(s,x) = X^\mu(0,x) + s  v^\mu(x),
\end{equation*}
where the fiducial embedding is taken to be $X^\mu(0,x) = \left(- x \sinh w , x \cosh w\right)$, the vector field $v^\mu(x) = \left( 
\cosh w , - \sinh w \right)$, and $w$ denotes the rapidity of the inertial frame moving at a relative speed $\beta$  defined as $\cosh w \ce 1/\sqrt{1-\beta^2}$. It then follows that
\begin{equation*}
    n_\nu(x) = \epsilon_{\nu\rho} \left. \frac{\partial {X}^\rho(x)}{\partial x} \right|_{\mathsf{X}  = \mathsf{X}_{v}(s) }  = (\cosh w , \sinh w ),
\end{equation*}
which is normal to each hypersurface of the foliation. For such a foliation, Eq.~\eqref{parmeterEvolution} reduces to the Sch\"{o}dinger equation 
\begin{equation*}
    i \frac{d}{dt} \ket{\psi_\phi[\mathsf{X}_{v}(t)]}  = \hat{H}_{{\rm flat}, \beta } \ket{\psi_\phi[\mathsf{X}_{v}(t)]} ,
    \label{Schrodingerflat}
\end{equation*}
where Eq.~\eqref{generator} simplifies to the Hamiltonian 
\begin{equation*}
    \hat{H}_{{\rm flat}, \beta } \ce \int d\Sigma(x) \, v^\mu(x)  n_\nu (x)  {\hat{T}^{\nu}}_{\ \mu}(x),
    \label{Hflat}
\end{equation*}
which generates an evolution in the group parameter $s=t$, interpreted as the proper time of an observer moving along the timelike congruence defined by the inertial foliation.

\subsubsection{Variation of the conditional hypersurface}

Instead of considering the conditional state $\ket{\psi_\phi[\mathsf{X}_{v}(s)]}$ to be a function of $s$, we may instead consider it to be a functional depending on the coordinate functions $X^\mu(x)=X_0^\mu(s) + s v^\mu(x)$, where $v^\mu(x)$ defines a one-dimensional subgroup $G_v \subset \mathrm{Diff}(\mathcal{M})$ characterizing a set of particular deformations of the fiducial hypersurface $\mathsf{X}_0$. Let us introduce the notation $\ket{\psi_\phi[{X}^\mu(x)]} \ce \ket{\psi_\phi[\mathsf{X}_{v}(s)]}$ to emphasize the functional dependence of the conditional state on the hypersurface $\mathsf{X}_{v}(s)$ through its coordinate functions ${X}^\mu(x)$. 

Now consider variations of the conditional state under deformations of the hypersurface $\mathsf{X}_v(s)$ by taking the functional derivative  with respect to the coordinate functions $X^\mu(x)$, while keeping $X_0^\mu(x)$ fixed for $s\in[s_0,s_1]$:
\begin{align}
    i \frac{\delta  \ket{\psi_\phi[{X}^\mu(x)]}}{\delta X^\mu(x)}  \!&=  i  \frac{\delta}{\delta X^\mu(x)} \braket{\mathsf{X}_{v}(s) | \Psi_{\rm phys}} \nn \\
    &= i \frac{\delta}{\delta X^\mu(x)} \bra{\mathsf{X}_0} e^{is\hat{\mathsf{\Pi}}[v]} \ket{\Psi_{\rm phys}} \nn \\
    &= i \bra{\mathsf{X}_0} \!\frac{\delta e^{i\! \int \!d\Sigma(x) \, \left( X^\mu(x) - X_0^\mu(x) \right) \hat{\Pi}_\mu(x)} }{\delta X^\mu(x)}  \! \ket{\Psi_{\rm phys}} \nn \\
    &= i \bra{\mathsf{X}_0} e^{is\hat{\mathsf{\Pi}}[v]} i \hat{\Pi}_\mu(x) \ket{\Psi_{\rm phys}} \nn \\
    &= - \bra{\mathsf{X}_{v}(s)} \hat{\Pi}_\mu(x) \ket{\Psi_{\rm phys}} \nn \\
    &=  \bra{\mathsf{X}_{v}(s)} \hat{h}_\mu(x) \ket{\Psi_{\rm phys}} \nn \\
    &= \hat{h}_\mu(x,X^\mu(x) ) \ket{\psi_\phi[{X}^\mu(x)]}.
    \label{preSchwingerTomanaga}
\end{align}
Let us define the surface variation as the normal projection of the functional derivative with respect to the coordinate functions \cite{Doplicher:2004gc}
\begin{equation*}
    \frac{\delta}{\delta \mathsf{X}(x)} \ce n^\mu(x) \frac{\delta}{\delta X^\mu(x)},
    \label{TomoSchwEq}
\end{equation*}
where $n^\mu(x)$ is the normal to $\mathsf{X}_{v}(s)$; this derivative characterizes normal deformations of the conditional wave functional. By contracting Eq.~\eqref{preSchwingerTomanaga} with $n^\mu(x)$, we recover the Tomonaga-Schwinger equation:
\begin{equation}
i\frac{\delta}{\delta \mathsf{X}(x)} \ket{\psi_\phi[{X}^\mu(x)]} = n^\mu(x)\hat{h}_\mu(x,X^\mu(x)) \ket{\psi_\phi[{X}^\mu(x)]}.
\label{SchwingerTomanaga}
\end{equation}

The recovery of the Tomonaga-Schwinger equation~equation~\cite{tomonaga_relativistically_1946,schwinger_quantum_1948,wakita1976integration,breuer_theory_2002} using the Page-Wootters formalism indicates its formal equivalence with  standard functional formulations of quantum field theory. One should note that the conformal anomaly potential $\hat{A}_\mu(x)$ is embedded in the Hamiltonian flux (for the $m=0$ case), underlying a redistribution of the energy flux over the embedding for hypersurfaces with non-vanishing extrinsic curvature. However, this once again does not affect the integrated version of the Tomonaga-Schwinger equation, as the anomaly vanishes for asymptotically flat embeddings. 

We recall that a functional Schr\"odinger equation has previously been derived in the context of PFT~\cite{*kuchar_dirac_1989,torre_quantum_1998,torre_functional_1999,varadarajan_dirac_2007,kaya_schrodinger_2022}; however, not on the reduced Hilbert space of conditional states of the scalar field as done here, but rather directly using the constraint and physical states. Employing the Page-Wootters formalism here permits us to instead identify such a functional Schr\"odinger equation with the Tomonaga-Schwinger equation, which similarly is formulated on the Hilbert space of the quantum field alone.

\subsubsection{Schr\"{o}dinger evolution along a one-parameter family of hypersurfaces}

Consider a one-parameter family $\mathcal{F} \ce \{ \mathsf{X}(t), \ \forall t \in [t_1,t_2] \}$ of spacelike hypersurfaces $\mathsf{X}(t)$ in Minkowski space (which may or may not constitute a foliation). Each hypersurface in the family is associated with a conditional state $\ket{\psi_\phi[\mathsf{X}(t)]}\ce \mathcal{R}_{\mathbf S} [\mathsf{X}(t)] \ket{\Psi_{\rm phys}}$, which evolves along the family according to: 
\begin{align} 
    i\f{d}{dt}\ket{\psi_\phi[\mathsf{X}(t)]}&=i\lim_{\varepsilon\to0}\f{\ket{\psi_\phi[\mathsf{X}(t+\varepsilon)]}-\ket{\psi_\phi[\mathsf{X}(t)]}}{\varepsilon} \nn  \\
    &=i\int d\Sigma(t,x)\, t^\mu(t,x) \f{\delta}{\delta {X}^\mu(x)}\ket{\psi_\phi[\mathsf{X}(t)]}\nn\\
    &=\hat H(t)\,\ket{\psi_\phi[\mathsf{X}(t)]}, \label{functderivative} 
\end{align}
where we have used Eq.~\eqref{preSchwingerTomanaga} to arrive at the last equality and defined the Hamiltonian
\begin{equation*}\label{eqHam}
\hat H(t)\ce\int d\Sigma  (t,x) \,t^\mu(t,x)\,\hat h_\mu(x,X^\mu(t,x)),
\end{equation*}
by projecting the energy-momentum density $\hat{h}_\mu(x,X^\mu(t,x))$ along the deformation vector field $t^\mu$ and integrating over the Cauchy surface $\Sigma$. In particular, we can decompose the Hamiltonian into normal and parallel deformation generators:
\begin{equation*}
\hat H(t)=\hat H_\bot(t)+\hat H_{\parallel}(t),
\end{equation*}
where
\begin{align*}
\hat H_\bot(t)&\ce \int d\Sigma(x) \, N(t)\,n^\mu(t,x)\,\hat h_\mu(x,X^\mu(t,x)),\nn\\
\hat H_\parallel(t)&\ce \int d\Sigma(x)\,N^\mu(t,x)\,\hat h_\mu(x,X^\mu(t,x)).\nn
\end{align*}
It is seen that Eq.~\eqref{functderivative}  constitutes a  generalized Schr\"odinger equation that accounts for normal deformations of the hypersurface $\mathsf{X}(t)$ generated by $\hat{H}_\perp (t)$, and spatial redistribution of points generated by $\hat{H}_\parallel(t)$ (i.e., spatial diffeomorphisms of $\mathsf{X}(t)$).  Moreover, the standard Schr\"odinger equation  in Minkowski space is recovered when restricting to inertial embeddings, such as $T(t,x)=t$ and $X(t,x)=x$. It then follows that $t^\mu(t,x)=n^\mu(t,x)=(1,0)$, so that
\begin{equation*}
    i\frac{d}{dt}\ket{\psi_\phi[\mathsf{X}(t)]} = \hat{H}_{{\rm flat},0} \ket{\psi_\phi[\mathsf{X}(t)]},
\end{equation*} 
where $\hat H_{{\rm flat},0} =\int d\Sigma(x)\, \hat T^0{}_0(x)$, is the standard Hamiltonian on a flat embedding and $\hat T^0{}_0(x)$ is the standard energy density in the employed coordinate frame. 

Integrating Eq.\eqref{functderivative}, we arrive at 
 \begin{equation}
    \ket{\psi_\phi[\mathsf{X}_2]} = \hat{U}_{\mathcal{F}}(t_2,t_1) \ket{\psi_\phi[\mathsf{X}_1]},
     \label{Usolution}
 \end{equation}
where
 \begin{equation}
     \hat{U}_{\mathcal{F}}(t_2,t_1) \ce  \mathcal{P} e^{ -i \!\int_{{t_1}}^{{t_2}} \!dt \, \hat H(t)} ,\label{evolgen}
\end{equation}
is formally a unitary operator that propagates the conditional state of the matter field from  $\mathsf{X}_1$ to $\mathsf{X}_2$, and $\mathcal{P}$ denotes the path-ordering operator in the foliation parameter $t$. Note that the Hamiltonian density satisfies the microcausality condition, $[\hat{h}^\mu(t,x),\hat{h}^\nu(t,x')] = 0$ for spacelike separated points labeled by the coordinates $(t,x)$ and $(t,x')$. This provides the integrability requirement for the integral in Eq.~\eqref{evolgen} to be well-defined and independent of the spacetime foliation of the region being integrated over~\cite{tomonaga_relativistically_1946,schwinger_quantum_1948,koba_1950,wakita1976integration,breuer_theory_2002}.
If the Hamiltonian generating $\hat{U}_{\mathcal{F}}(t_2,t_1)$ is constructed by smearing the Hamiltonian flux with a deformation vector field associated with a complete spacelike foliation, then it is a complete vector field in Minkowski space $\mathcal{M}$. The deformation vector fields lifts to a complete vector field on the classical scalar field configuration space $\mathcal{C}_\phi$ (see Sec.~\ref{sec: DiracPFT}) because the deformation vector field generates a gauge diffeomorphism, which cannot change the $C^2$ nature of a classical field configuration, nor its asymptotic drop-off property, given that also gauge diffeomorphisms have to vanish asymptotically. 
In fact, this should extend to the quantum configuration space $\mathcal{S}'$, where the action of the Lie derivative on distributions can be understood via partial integration and the properties of the gauge diffeomorphisms should ensure the preservation of the properties of the permitted set of distributions. Making this rigorous is somewhat challenging; however, it suggests that van Hove's theorem in quantum  mechanics~\cite{VanHove,Gotay:1996yx,Gotay:1998aw} could be extrapolated to field theory, in which case it would imply that the Hamiltonian is formally self-adjoint. 

\begin{figure}[t]
    \centering
    \includegraphics[width=0.48\textwidth]{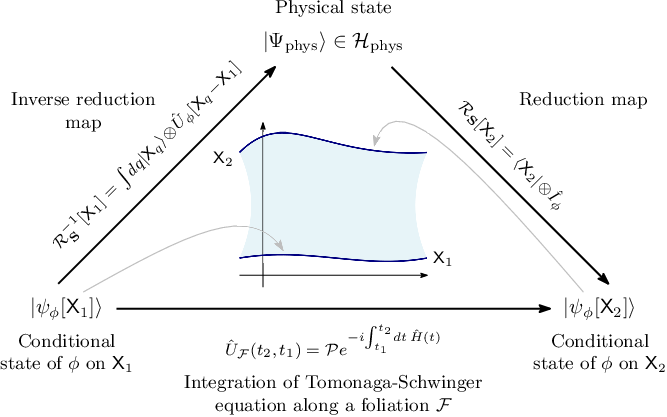}
    \caption{The composition of the inverse reduction map relative to $\mathsf{X}_1$ and reduction map relative to $\mathsf{X}_2$ (upper path), which maps the conditional state $\ket{\psi_{\phi}[\mathsf{X}_1]}$ to  $\ket{\psi_{\phi}[\mathsf{X}_2]}$, is shown in Eq.~\eqref{FoliationIndependence} to be equivalent to integrating the Tomonaga-Schwinger equation along an arbitrary spacelike foliation of Minkowski space  connecting $\mathsf{X}_1$ and $\mathsf{X}_2$ (lower path). Foliation-independence of the evolution of the conditional state from $\mathsf{X}_1$ and $\mathsf{X}_2$ arises in the former as a consequence of the diffeomorphisim invariance of the physical states from which the conditional states of the scalar field are constructed, while in the latter it arises as a consequence of the microcausality condition satisfied by the stress-energy tensor of the scaler field.}
\label{fig:IntegrationOfTS}
\end{figure}

Alternatively, we may evolve $\ket{\psi_\phi[\mathsf{X}_1]}$ to $\ket{\psi_\phi[\mathsf{X}_2]}$ by first applying the inverse of the reduction map evaluated at $\mathsf{X}_1$ and then the reduction map evaluated at $\mathsf{X}_2$, yielding:
\begin{align}
        \ket{\psi_\phi[\mathsf{X}_{2}]} &= \mathcal{R}_{\mathbf{S}}[\mathsf{X}_{2}] \circ \mathcal{R}_{\mathbf{S}}^{-1}[\mathsf{X}_1]   \ket{\psi_\phi[\mathsf{X}_1]} \nn \\  &=\bra{\mathsf{X}_{2}} \otimes \hat{I}_\phi\int\mathcal{D}q\,\ket{\mathsf{X}_{q}} \otimes\hat{U}_\phi[\mathsf{X}_{q}-\mathsf{X}_{1}] \ket{\psi[\mathsf{X}_{1}]} \nn \\
        &=\hat{U}_\phi[\mathsf{X}_{2}-\mathsf{X}_{1}] \ket{\psi[\mathsf{X}_{1}]}. \label{eq:embstatechange}
    \end{align}
Assuming the family of hypersurfaces constitutes a foliation, applying Eq.~\eqref{eq:embstatechange}  between successive leaves, as is done in Appendix~\ref{app: unitaryeqiv}, together with the fact that $\mathcal{R}_{\mathbf{S}}^{-1}[\mathsf{X}_q] \circ \mathcal{R}_{\mathbf{S}}[\mathsf{X}_q] = \hat{I}_{\rm phys}$, which stems from the gauge invariance of physical states imposed by Eq.~\eqref{QuantumConstraint}, it follows that we must have
\begin{equation}
    \ket{\psi_\phi[\mathsf{X}_{2}]}=  \hat{U}_{\mathcal{F}}(t_2,t_1)\ket{\psi_\phi[\mathsf{X}_{1}]} = \hat{U}_\phi[\mathsf{X}_{2}-\mathsf{X}_{1}] \ket{\psi_\phi[\mathsf{X}_{1}]}.
    \label{FoliationIndependence}
\end{equation}
This shows that the evolution between the two hypersurfaces is formally unitary and foliation independent  since neither the reduction map nor its inverse depends on $\mathcal{F}$. This is a different way of demonstrating the foliation-independence of the Tomonaga-Schwinger equation that does not use the microcausality condition but instead uses the fact that physical states are invariant under diffeomorphisms, which is a gauge symmetry of PFT. See Fig.~\ref{fig:IntegrationOfTS} for more detail. This is an instantiation of the fact that foliation-independence and diffeomorphism invariance are deeply intertwined in canonical formulations of generally covariant theories, e.g.\ see \cite{kieferQuantumGravity2012,Hojman:1976vp,Kuchar:1976yx,Kuchar:1976yy, thiemannModernCanonicalQuantum2008}.

\subsection{Changes of embedding configuration}

Having constructed the conditional states $\ket{\psi_\phi[\mathsf{X}_q]}$ of the scalar field relative to an embedding $\mathsf{X}_q$, we now consider how the conditional state changes under a change of embedding, $\mathsf{X}_q \to \mathsf{X}_{q'}$. Since the embedding fields constitute a dynamical reference frame, each configuration can be regarded as a reference frame orientation. The action of a generic change in frame configuration, $\mathsf{X}_q \to \mathsf{X}_{q'}$, on the conditional state  $\ket{\psi_\phi[\mathsf{X}_q]}$ is given in terms of the reduction map and its inverse, and is seen to induce a unitary  map on conditional states:
\begin{equation}
\begin{split}
     \ket{\psi_\phi[\mathsf{X}_{q'}]} &= \mathcal{R}_{\mathbf{S}}[\mathsf{X}_{q'}] \circ \mathcal{R}_{\mathbf{S}}^{-1}[\mathsf{X}_q]   \ket{\psi_\phi[\mathsf{X}_q]}  \\
     &= \hat{U}_\phi[\mathsf{X}_{q'}-\mathsf{X}_{q}] \ket{\psi_\phi[\mathsf{X}_q]}.\label{eq:orientchangeNew}
     \end{split}
\end{equation}

As depicted in Fig.~\ref{fig:IntegrationOfTS}, this map is used to transform conditional states between two embedding configurations associated with different leaves of a foliation. However, the map in Eq.~\eqref{eq:orientchangeNew} is more general as it maps the conditional state between two embedding configurations that are not necessarily part of a foliation. For example, consider two embedding configurations related by a finite Lorentz boost $\Lambda$. The coordinate functions characterizing these embeddings are related by $X^\mu_{\Lambda q}(x)={\Lambda^\mu}_\nu X^{\nu}_q(x)$. Inserting this in Eq.~\eqref{eq:orientchangeNew}, we arrive at:
\begin{equation*}
    \begin{split}
       \ket{\psi_\phi[\mathsf{X}_{\Lambda q}]}&=\hat{U}_{\phi}[\mathsf{X}_{\Lambda q}-\mathsf{X}_q]\ket{\psi_\phi [\mathsf{X}_q]}\\
       &=e^{-i\int d\Sigma(x) \,({\Lambda^\mu}_\nu-{\delta^\mu}_\nu) X_q^\nu(x)   \hat{h}_\mu(x)} \ket{\psi_\phi [\mathsf{X}_q]}.
    \end{split}
\end{equation*}
Furthermore, covariance requires that field observables $\hat{A}_\phi(q)$ transform between hypersurfaces $q$ and $\Lambda q$ as
\begin{equation*}
        \hat{A}_\phi(\Lambda q)=\hat{U}_\phi[\mathsf{X}_{\Lambda q}-\mathsf{X}_q]\hat{A}_\phi(q)\hat{U}^\dagger_\phi[\mathsf{X}_{\Lambda q}-\mathsf{X}_q].
    \end{equation*}

We may also consider how the conditional state transforms under a transformation from an inertial to a uniformly accelerating frame~\cite{fulton_rohrlich_witten_1962,wood_papini_cai_1989}.  In particular, if $b^\mu\ce\left(0,\frac{1}{\alpha} \right)$ is a constant vector, the embedding configuration obtained through the finite special conformal transformation $X^\mu_{q'}(x)=\frac{X^\mu_{q}(x)-b^\mu X^2_{q}(x)}{\beta(x)}$ experiences a uniform acceleration $\alpha$ relative to $X_q$, where $\beta(x)\ce 2X^\sigma(x) b_\sigma- X_q^\sigma(x){X_q}_\sigma(x)/\alpha^2$. Using Eq.~\eqref{eq:orientchangeNew}, the conditional state of the field transforms as 
    \begin{equation*}
    \begin{split}
        \ket{\psi_\phi[\mathsf{X}_{q'}]}
        &=\hat{U}_{\phi}[\mathsf{X}_{q'}-\mathsf{X}_q]\ket{\psi_\phi [\mathsf{X}_q]}\\
        &=e^{-i\int d\Sigma(x) \, \left[\frac{\beta(x) X_q^\mu(x)-b^\mu X_q^2(x)}{1-\beta(x)}\right]  \hat{h}_\mu(x)}\ket{\psi_\phi [\mathsf{X}_q]}.
    \end{split}
    \end{equation*}
Note that this operator may also a non-trivial domain, as $X_{q'}^\mu(x)$ defined via the acceleration above is not in general guaranteed to be an embedding itself.

\subsection{Construction of relational Dirac observables}
\label{Sec:DiracObservables}

We may construct gauge-invariant observables directly on the physical Hilbert space that encode relations between the matter field and embedding fields; such observables are known as relational Dirac observables \cite{rovelliQuantumGravity2004,thiemannModernCanonicalQuantum2008,dittrichPartialCompleteObservables2006a,dittrichPartialCompleteObservables2007a,Goeller:2022rsx}.  Consider an observable on the field Hilbert space
\begin{equation*}
    \hat{A}_0 \ce  \int d\Sigma_0(x) \, A[ \hat{\phi}(x),\hat{\pi}(x)] \in \mathcal{B}(\mathcal{H}_\phi).
\end{equation*}
Using the reduction map and its inverse in Eqs.~\eqref{defReduction} and \eqref{defReductionInverse}, we can construct a relational Dirac observable $\hat{F}_{A_0}[\mathsf{X}_q]$ on the physical Hilbert space~\cite{hohn_trinity_2021}, such that
\begin{equation*}
\bra{\psi_\phi[\mathsf{X}_q]}     \hat{A}_0  \ket{\psi_\phi[\mathsf{X}_q]} = \bra{\Psi_{\rm phys}}     \hat{F}_{A_0}[\mathsf{X}_q]  \ket{\Psi_{\rm phys}}_{\rm phys},
\label{ExpectationvalueEquivelence}
\end{equation*}
where the subscript `phys' denotes the physical inner product as opposed to the kinematical inner product (no subscript), and
\begin{align}
    \hat{F}_{A_0}[\mathsf{X}_q] &\ce \mathcal{R}^{-1}_{\bf S}[\mathsf{X}_q]     \hat{A}_0  \mathcal{R}_{\bf S}[\mathsf{X}_q] \label{RelationDiracDef} \\
    &= \int \mathcal{D}q' \, \hat{U}_{\mathsf{X}}[\mathsf{X}_{q'} \!-\! \mathsf{X}_{q}] \ket{\mathsf{X}_q} \! \bra{\mathsf{X}_q} \otimes \hat{U}_{\phi}[\mathsf{X}_{q'} \!- \!\mathsf{X}_{q}]      \hat{A}_0.\nn
\end{align}
As shown  in Appendix~\ref{DiracObservablesApeendix}, $\hat{F}_{A_0}[\mathsf{X}_q]$ commutes with the constraint $\hat{\mathsf{C}}[g]$ on physical states,
\begin{equation*}
    \big[ \hat{\mathsf{C}}[g], \hat{F}_{A_0}[\mathsf{X}_q] \big] \ket{\Psi_{\rm phys}} = 0,
\end{equation*}
and thus constitutes a gauge invariant relational Dirac observable. $\hat F_{A_0}[\mathsf{X}_q]$ encodes the outcome of  the observable $\hat A_0$, conditioned on the embedding being in configuration $\mathsf{X}_q$. 

The dynamics of the matter field $\phi$ relative to the embedding fields $X^\mu$ along a spacelike foliation $\mathcal{F} = \{\mathsf{X}(t), [t_1,t_2]\}$ is encoded in the one-parameter family of relational Dirac observables $F_{A_0}[\mathsf{X}(t)]$ acting on $\mathcal{H}_{\rm phys}$. This should be contrasted with how the same dynamics in the Page-Wootters formalism is encoded in the integration of the Tomonaga-equation on $\mathcal{H}_\phi$ along $\mathcal{F}$ as in Eq.~\eqref{evolgen} and Fig.~\ref{fig:IntegrationOfTS}.

\subsection{Quantum deparametrization to a relational Heisenberg picture}

Conditioning physical states on embedding states yielded a relational functional Schr\"odinger picture for the free scalar field that formally encompasses the standard wave functional formulation in Minkowski space. As a next step, we apply a quantum deparametrization procedure related to the method developed in Refs.~\cite{hoehnHowSwitchRelational2018,Hoehn:2019,hohn_trinity_2021,hohn_equivalence_2021,de_la_hamette_perspective-neutral_2021} to give a unitarily equivalent, functional Heisenberg picture for the scalar field. Such a procedure is modeled on the classical analog of deparameterizing a classical reparametrization-invariant theory.

Let us consider the following deparametrization map relative to a fiducial hypersurface $\mathsf{X}_0$\footnote{The analog in \cite{hoehnHowSwitchRelational2018,Hoehn:2019,hohn_trinity_2021,hohn_equivalence_2021,de_la_hamette_perspective-neutral_2021} trivializes the constraints to the reference system, thereby effectively disentangling (relative to the kinematical tensor product structure) the latter and the degrees of freedom to be described relative to it. In the present case, the map does not achieve this for arbitrarily smeared constraints.}
\begin{equation}
\mathcal{T}_{\mathsf{X}_0} \ce \int \mathcal{D}q \, \ket{\mathsf{X}_q}  \! \bra{\mathsf{X}_{q}} \otimes\hat{U}_{\phi}[\mathsf{X}_{0} - \mathsf{X}_{q}],
\end{equation}
with inverse
\begin{equation}
 \mathcal{T}^{-1}_{\mathsf{X}_0} = \int \mathcal{D}q \, \ket{\mathsf{X}_q} \!\bra{\mathsf{X}_q} \otimes \hat{U}_{\phi}^\dagger[\mathsf{X}_{0} - \mathsf{X}_{q}] , 
\end{equation}
such that $\mathcal{T}^{-1}_{\mathsf{X}_0} \circ \mathcal{T}_{\mathsf{X}_0}  = \hat{I}_{\rm phys}$.

We define the Heisenberg reduction map as 
\begin{align*}
    \mathcal{R}_{\mathbf H}: \mathcal{H}_{\rm phys} &\rightarrow \mathcal{H}_{\mathsf{X}_q}; \nn \\ \ \ket{\Psi_{\rm phys}} &\mapsto\mathcal{R}_{\mathbf H} \ket{\Psi_{\rm phys}},
\end{align*}
where
\begin{equation}
\mathcal{R}_{\mathbf H} \ce \bra{\mathsf{X}_q}   \mathcal{T}_{\mathsf{X}_0} =  \hat{U}_{\phi}[\mathsf{X}_{0} - \mathsf{X}_{q}]  \mathcal{R}_{\mathbf S}[\mathsf{X}_q] = \mathcal{R}_{\mathbf S} \left[ \mathsf{X}_0 \right],
\label{HisenbergReducation}
\end{equation}
with inverse 
\begin{align*}
\mathcal{R}_{\mathbf H}^{-1}  &\ce   \mathcal{T}^{-1}_{\mathsf{X}_0} \!\int \mathcal{D}q\, \ket{\mathsf{X}_q}  \otimes \hat{I}_\phi =\! \int \mathcal{D}q \, \ket{\mathsf{X}_q} \otimes \hat{U}_{\phi}^\dagger[\mathsf{X}_{0} - \mathsf{X}_{q}] .
\end{align*}
Given a physical state $\ket{\Psi_{\rm phys}}$, the Heisenberg reduction map and its inverse can be used to construct a relational Heisenberg picture.  Analogous to the conditional state defined in Eq.~\eqref{ConditionalWaveFunctional}, we define a conditional Heisenberg state relative to a fiducial embedding $\mathsf{X}_0$ as
\begin{equation*}
    \mathcal{R}_{\mathbf H} \ket{\Psi_{\rm phys}} = \mathcal{R}_{\mathbf S}[\mathsf{X}_0] \ket{\Psi_{\rm phys}} = \ket{\psi_{\phi}[\mathsf{X}_0]}.
\end{equation*}
In addition, the relational Dirac observables defined in Eq.~\eqref{RelationDiracDef} can be mapped to Heisenberg picture observables (cf., Theorem 5 of Ref.~\cite{hohn_trinity_2021})
\begin{align*}
     \hat{A}_\phi[\mathsf{X}_q] &\ce \mathcal{R}_{\mathbf H}  \hat{F}_{A_0}[\mathsf{X}_q] \mathcal{R}_{\mathbf H}^{-1} \nn \\
     &= \hat{U}_{\phi}[\mathsf{X}_{0}  - \mathsf{X}_{q}]      \hat{A}_0\, 
 \hat{U}_{\phi}^\dag[\mathsf{X}_{0} - \mathsf{X}_{q}],
\end{align*}
 which satisfies the Heisenberg functional equation of motion 
\begin{equation}
     \frac{\delta }{\delta X^\mu_q(x)}\hat{A}_\phi[\mathsf{X}_q]=i\left[h_\mu[\mathsf{X}_q, x],\hat{A}_\phi\right],
     \label{Hisenberg}
\end{equation}
as we show  in Appendix~\ref{app: calculations}.
Thus, together with the Tomonaga-Schwinger equation in Eq.~\eqref{SchwingerTomanaga}, the relation between the Heisenberg and Schr\"{o}dinger reduction maps in Eq.~\eqref{HisenbergReducation}, and the definition of the Dirac observables in Eq.~\eqref{RelationDiracDef}, it is seen that the description of a PFT in terms of the Page-Wootters formalism developed in Sec.~\ref{PWconstruction}, the quantum deparameterization procedure introduced immediately above, and the relational Dirac observable prescription presented in Sec.~\ref{Sec:DiracObservables} yield relational formalisms that are unitarily equivalent. Thus, as in the case of non-interacting mechanical systems with vanishing Hamiltonian constraints~\cite{hohn_trinity_2021,hohn_equivalence_2021}, there exists a field-theoretic trinity of relational quantum dynamics for PFT.

\section{Quantum embedding transformations}
\label{Sec: QETransformation}

Analogous to the usual interpretation of the conditional state in the Page-Wootters formalism, the conditional state $\ket{\psi_\phi[\mathsf{X}]}$ is a functional describing a matter field $\phi$ relative to a hypersurface associated with the embedding $\mathsf{X}$. Operationally, the embedding fields serve as abstractions of a reference frame consisting of the rods and clocks of a congruence of observers.

Given this interpretation, an important task is to examine how the conditional wave functional $\ket{\psi_{\phi}[\mathsf{X}]}$ transforms under a change of embedding $\mathsf{X}$, and in doing so, recover classical frame transformations in a certain approximation. Moreover, the formalism introduced above treats embeddings as genuine quantum degrees of freedom associated with a Hilbert space, allowing us to go beyond classical frame transformations and examine the field from the perspective of a nonclassical embedding. Such a generalization is in the spirit of Ref.~\cite{giacominiQuantumMechanicsCovariance2019} in which spatial superposition and entanglement were interchangeable under an analogous quantum frame transformation. Moreover, such quantum frame transformations provide the foundation on which to build a quantum theory of general covariance~\cite{Hoehn:2019,hoehnHowSwitchRelational2018,vanrietveldeChangePerspectiveSwitching2020,hamette_quantum_2020,hohn_trinity_2021,hohn_equivalence_2021, de_la_hamette_perspective-neutral_2021,vanrietveldeSwitchingQuantumReference2018}, which is expected to play an important role in a quantum theory of gravity.

\subsection{Changes of embedding fields}

In the remainder, we shall now assume that there exist two dynamically distinct embedding fields, either of which we can employ as a reference frame. For example, one of them might be given by some additional matter fields, though we shall not specify how this effects the Hamiltonian constraint. In this way, while being more formal, we are also more general.

Consider then two embedding Hilbert spaces $\mathcal{H}_A$ and $\mathcal{H}_B$ with their respective embedding states denoted as $\ket{\mathsf{X}_A}$ and $\ket{\mathsf{Y}_B}$. Suppose the conditional wave functional relative to $A$ is $\ket{\psi_\phi[\mathsf{X}_A]}$, and the state of embedding field $B$ is 
\begin{equation*}
    \ket{\psi_{B|A} [\mathsf{X}_A]} = \int \mathcal{D} q \, \psi_{B}[\mathsf{Y}_q,\mathsf{X}_A] \ket{\mathsf{Y}_{q}},
\end{equation*} 
so that their joint state is 
\begin{equation}
\ket{\psi_{B,\phi }[\mathsf{X}_A]}= \ket{\psi_{B|A}} \ket{\psi_\phi[\mathsf{X}_A]}. 
\label{BrelativeA}
\end{equation}

The change of embedding map $\Lambda_{A \to B}: \mathcal{H}_B \otimes \mathcal{H}_\phi \to \mathcal{H}_A \otimes \mathcal{H}_\phi$ transforms a conditional state relative to embedding $A$ to the conditional state relative to embedding~$B$
\begin{align*}
    \Lambda_{A \to B} &\ce \mathcal{R}_{\mathbf{S}}[\mathsf{Y}_B] \circ \mathcal{R}_{\mathbf{S}}^{-1}[\mathsf{X}_A] \\
    &=\int \mathcal{D}q \, \ket{\mathsf{X}_q} \otimes \bra{\mathsf{Y}_B}  \hat{U}_{\mathsf{Y}}[\mathsf{X}_q- \mathsf{X}_A] \otimes \hat{U}_\phi[\mathsf{X}_q- \mathsf{X}_A]. \nn 
\end{align*}

Transforming the conditional state in Eq.~\eqref{BrelativeA}, defined as the state of $B$ and $\phi$ relative to $A$, to a conditional state of $A$ and $\phi$ relative to $B$:
\begin{align*}
  \ket{\psi_{A,\phi}[\mathsf{Y}_B]} &= \Lambda_{A \to B} \ket{\psi_{B,\phi}[\mathsf{X}_A]} \nn \\
       &= \int \mathcal{D}q  \, \psi_{B|A}\left[\mathsf{Y}_B- \mathsf{Y}_{q} \right] \ket{\mathsf{X}_q}   \ket{\psi_\phi[\mathsf{X}_q]}.
\end{align*}
It is seen that even though the matter field $\phi$ and embedding fields associated with $B$ are seperable relative to $A$, from the perspective of $B$, they are entangled. This is analogous to the frame-dependence of superposition and entanglement exhibited in Ref.~\cite{giacominiQuantumMechanicsCovariance2019,ali_ahmad_quantum_2022}.

\subsection{Particle creation due to reference frame coherence}

Having defined the change of embedding map $\Lambda_{A \to B}$, its action can transform observables between foliations of spacetime associated with distinct embedding frame fields. In particular, consider two foliations $\mathcal{F}_A = \{\mathsf{X}_A(\tau)\}$ and $\mathcal{F}_B = \{\mathsf{X}_B(t)\}$ of $\mathcal{M}$, each associated with a timelike killing vector field, $\partial_\tau$ and $\partial_t$, respectively. Suppose the field is expanded in creation and annihilation operators  along $\mathcal{F}_A$ as 
\begin{equation*}
 \hat{\phi}(\tau,x) =\int dk \, \left( u_k(\tau,x)  \hat{a}_k + u_k^*(\tau,x)]   \hat{a}_k^\dagger  \right),
\end{equation*}
and along $\mathcal{F}_B$ as
\begin{equation*}
 \hat{\phi}(t,x) = \int dk \, \left( v_k(t,x) \hat{b}_k + v_k^*(t,x)   \hat{b}_k^\dagger  \right),
\end{equation*}
where $u_k(\tau,x)$ and $u_k(\tau,x)$, together with their complex conjugate, form two complete sets of orthonormal solutions to the classical field equation.

The number operator $\hat{N}_{k,A} \ce a_k^\dagger a_k$, which counts the number of particles relative to $\mathcal{F}_A$, can be expressed relative to $\mathcal{F}_B$, as follows:
\begin{align*}
\hat{N}_{k,A} &\mapsto \Lambda_{A \to B} \big(\hat{I}_B \otimes   \hat{N}_{k,A} \big)\Lambda_{A \to B}^\dagger \nn \\
&= \int \mathcal{D} q  \ket{\mathsf{X}_q} \!\bra{\mathsf{X}_{q}} \otimes  \hat{N}_{k,q},   
\end{align*}
where $\hat{N}_{k,q} \ce \hat{c}_{k,q}^\dagger\hat{c}_{k,q}$ is the number operator on the hypersurface $\mathsf{X}_q$, and $\hat{c}_{k,q}$ and $\hat{c}_{k,q}^\dagger$ are the associated creation and annihilation operators with mode functions $w_i(\tilde{t},x)$, where $\tilde{t}$ is a time coordinate along a foliation between $\mathsf{X}_A$ and $\mathsf{X}_q$. In particular, observe that the transformation maps a local operator into a non-local one. We note in passing that the transformed number operator has the same form as the encoding map that appears in the context of quantum communication~\cite{bartlettReferenceFramesSuperselection2007,bartlettQuantumCommunicationUsing2009,smithQuantumReferenceFrames2016,smithCommunicatingSharedReference2018}, as a quantization of Dirac observables using covariant POVMs~\cite{hohn_trinity_2021,chataignier_construction_2020,chataignier_relational_2021,de_la_hamette_perspective-neutral_2021}, and as a relativization map in the context of Refs.~\cite{loveridgeSymmetryReferenceFrames2018,carette_operational_2023,glowacki_quantum_2023}. The operators $\hat{c}_{k,q}$ can be expressed as a linear combination of the creation and annihilation operators $\hat{b}_k$ and $\hat{b}_k^\dagger$ via the Bogoliubov transformation
\begin{equation*}
    \hat{c}_{k,q} = \int dj \left(\alpha_{jk}(q) \hat{b}_j + \beta^*_{jk}(q) \hat{b}_j^\dagger\right),
\end{equation*}
where $\alpha_{jk}(q) \ce \left(v_j,w_k \right)$ and $\beta_{jk}(q) \ce -\left( v_j,w_k^*\right)$ are Bogoliubov coefficients and $(\, , \, )$ is the usual Klein-Gordon inner product on $\mathsf{X}_q$.

Now suppose that relative to $\mathcal{F}_B$ the scalar field is in its vacuum state $\ket{0_{\phi|B}}$, and the state of embedding $A$ is $\ket{\psi_{A|B}}$. Then, the expected number of particles relative to $A$ is
\begin{align}
   \bra{\psi_{A|B}} \bra{0_{\phi|B}}  &\hat{N}_{k,A} \ket{\psi_{A|B}} \ket{0_{\phi|B}} \nn \\
      &= \int \mathcal{D} q  \abs{\psi_{A|B} \left[ \mathsf{X}_q\right]}^2   \int dj\, \abs{\beta_{jk}(q)}^2.
      \label{genBogo}
\end{align}
Supposing that $|\psi_{A|B} \left[ \mathsf{X}_q\right]|^2 $ has support on $\mathsf{X}_q$ that have non-vanishing $\beta_{jk}(q)$, the vacuum state of $B$ will be seen to have particles relative to $A$. A `classical' Bogoliubov transformation is recovered when $|\psi_{A|B} \left[ \mathsf{X}_q\right]|^2 \to \delta \left[\mathsf{X}_q - \mathsf{X}_B\right]$ becomes sharply localized around $\mathsf{X}_B$.

A possible interpretation of Eq.~\eqref{genBogo} is that standard treatments of quantum field theory miscount the number of field quanta by assuming an infinitely precise localization of the laboratory frame. In contrast, Eq.~\eqref{genBogo} weighs the expected particle number for each embedding by the distribution $\abs{\psi_{A|B} \left[ \mathsf{X}_q\right]}^2$. This particle creation effect is a field-theoretic analogue of the transformation between superposition and entanglement that occurs in general quantum reference frame changes discussed in Ref.~\cite{giacominiQuantumMechanicsCovariance2019,ali_ahmad_quantum_2022}, and the interplay between clock state localization and temporally localized dynamics discussed~\cite{ruizQuantumClocksTemporal2020,hohn_trinity_2021}. We note that related, though qualitatively different, particle creation effects have been discussed in the context of superpositions of accelerated observers~\cite{barbado_unruh_2020} and superpositions of spacetime structure~\cite{foo_quantum_2023, foo_quantum_2022,Kabel2022spacetimesuperpos}.

\section{Discussion and outlook}
\label{sec: discussion}

Beginning with the Dirac quantization of PFT, we have developed a field-theoretic generalization of the Page-Wootters approach to relational dynamics. We began by formally constructing the kinematical Hilbert spaces of the scalar matter field and embedding fields. We then introduced group coherent states of the embedding fields relative to subgroups of the diffeomorphisms of Minkowski space, and used them to construct states of the scalar field relative to hypersurface configurations. Exploiting the group properties of these coherent states, we showed that the conditional state of the scalar field satisfies the Tomonaga-Schwinger equation, and thus demonstrated that the field-theoretic generalization of the Page-Wootters formally coincides with standard functional formulations of quantum field theory. As depicted in Fig.~\ref{fig:IntegrationOfTS}, what is notable in this framework is the appearance of a dual picture connecting the foliation-independence of the conditional state's evolution between two spacelike hypersurfaces emerging as a consequence of the invertibility of the Page-Wootters reduction maps, and ultimately the gauge-invariance of physical states. Alternatively, foliation-independence of the integrated Tomonaga-Schwinger is usually derived as a consequence of the microcauslity condition satisfied by the scalar field's stress-energy tensor. 

In addition, we constructed relational Dirac observables corresponding to field observables relative to configurations of the embedding fields and a quantum deparameterization of the physical Hilbert space, which led to a relational Heisenberg picture~\cite{hoehnHowSwitchRelational2018,Hoehn:2019,hohn_trinity_2021,hohn_equivalence_2021,hohn_trinity_2021,hohn_equivalence_2021}. Both of these relational frameworks were shown to be unitarily equivalent to the Page-Wootters formalism, extending the equivalence established in Refs.~\cite{hohn_trinity_2021,hohn_equivalence_2021,de_la_hamette_perspective-neutral_2021}. Finally, we introduced a field-theoretic extension of quantum reference frame changes~\cite{giacominiQuantumMechanicsCovariance2019}, which can be used to transform the number operator between different quantum reference frames characterized by a quantum state of the embedding fields. In doing so, we showed that the particle content of a field depends on the quantum state of the embedding it is being described relative to, highlighting a consequence of treating embeddings as quantum reference frames.

Natural generalizations of the above analysis are possible, though they do not come without difficulty. It would be natural to extend the Page-Wootters formalism to higher spacetime dimensions; however, the canonical quantization of such a PFT leads to difficulties in constructing unitary evolution along arbitrary foliations~\cite{torre_quantum_1998,torre_functional_1999}. Nevertheless, extensions to higher spacetime dimensions have been developed either through the use of loop quantization techniques~\cite{varadarajan_dirac_2007} or through algebraic methods~\cite{torre_functional_1999}, the latter of which implements evolution between spacelike hypersurfaces via a map that has a similar compositional structure as the evolution map depicted in Fig.~\ref{fig:IntegrationOfTS}.  Alternatively, one could imagine generalizing the framework by replacing the Minkowski background considered here with a curved spacetime, though the simple difference of vector fields used to connect points on different hypersurfaces would have to be suitably generalized (e.g., Fermi-Walker transport). Instead of modifying the background spacetime, another avenue to explore could be replacing the `ideal' embedding fields with physical matter that possesses its own independent dynamics. Such a generalization is directly relevant to the canonical quantization program given that the canonical quantization of general relativity with dust is deparamterizatble like PFT~\cite{brown_dust_1995,Tambornino:2011vg,Goeller:2022rsx,Giesel:2007wi,Giesel:2007wn,Husain:2011tk,Rovelli:1990ph} (see also Ref.~\cite{kaya_schrodinger_2022}). In principle, a Page-Wootters formulation of geometrodynamics may be possible, even if such a goal is ambitious. Furthermore, it would be interesting to explore connections with the recently proposed geometric event-based relativistic quantum mechanics~\cite{giovannetti_geometric_2022},  which is based on a similar conditioning procedure that was employed in this work, as well as the ``second quantization'' of the Page-Wootters formalism proposed in \cite{Diaz:2020dfe,Diaz:2021snw}.

We further mention that phase space extensions by embedding fields also play a crucial role in the recent discussions on finite subregions and edge modes in gravity \cite{Donnelly:2016auv,Speranza:2017gxd,Speranza:2019hkr,Freidel:2020xyx,Freidel:2023bnj,Ciambelli:2021nmv,Carrozza:2022xut,Kabel:2023jve}. In that context, these embedding fields rather define the spatial boundaries of the subregion in a relational manner. Again, a conditional state formulation may also be applicable to such a setting. 

Aside from the dynamical model provided by PFT, our analysis relies on coherent states of the embedding field introduced in Eq.~\eqref{DefCoherntStates}, which correspond to classical configurations characterizing hypersurfaces in Minkowski space. Group coherent states play an important role in many applications of quantum theory~\cite{perelomovGeneralizedCoherentStates1986}, and often serve to indicate the orientation of reference frame~\cite{buschQuantumTheoryMeasurement1991,*buschOperationalQuantumPhysics1995,*buschQuantumMeasurement2016, bartlettReferenceFramesSuperselection2007, smithQuantizingTimeInteracting2017,smith_quantum_2020,hohn_trinity_2021, hohn_equivalence_2021,hamette_quantum_2020,de_la_hamette_perspective-neutral_2021,loveridgeSymmetryReferenceFrames2018,carette_operational_2023,glowacki_quantum_2023,chataignier_construction_2020,chataignier_relational_2021}.  However, the literature has focused on coherent states associated with finite-dimensional Lie groups, whereas the coherent states of the embedding fields considered above are associated with subgroups of the infinite-dimensional Lie group of spacetime diffeomorphisms, and further study of their mathematical structure is of interest~\cite{twareque_ali_quantization_1991}.

\begin{acknowledgments}

We thank S.~Ali Ahmad for comments on an early version of this manuscript. PH is grateful for the hospitality of the high-energy physics group at EPF Lausanne, where the final stages of this work were carried out. This work was supported by funding from Okinawa Institute of Science and Technology Graduate University and initially by an ``It-from-Qubit'' Fellowship of the Simons Foundation (awarded to PH). This project/publication was also made possible through the support of the ID\# 62312 grant from the John Templeton Foundation, as part of the \href{https://www.templeton.org/grant/the-quantum-information-structure-of-spacetime-qiss-second-phase}{\textit{`The Quantum Information Structure of Spacetime'} Project (QISS)}.~The opinions expressed in this project/publication are those of the author(s) and do not necessarily reflect the views of the John Templeton Foundation. ARHS is grateful for financial support through a Summer Research Grant awarded by Saint Anselm College.
\end{acknowledgments}

\appendix
\onecolumngrid

\section{Explicit calculations}
\label{app: calculations}
In this appendix, we collect explicit calculations leading to some of the results in the main body. In particular, we describe the connection between eigenstates of the embedding operator and coherent states, and we prove the commutation of lifted Dirac observables with the constraint operator.

\subsection{Generalized eigenstates of the embedding operator as group coherent states}
\label{ProofOfTheorems}

Here we show that generalized eigenstates $\ket{\mathsf{X}_v(s)}$ of the embedding operator $\hat{X}^\mu(x)$ are coherent states associated with one-dimensional subgroups $G_v \subset G$ of the group $G \ce \text{Diff} ( \mathcal{M})$ with eigenfunctions corresponding to the coordinate functions of the associated embedding $\mathsf{X}_v(s):\Sigma \to \mathcal{M}$, where $s\in[s_0,s_1]$.

Consider the coherent state system relative to the group $G_v$ introduced in Eq.~\eqref{DefCoherntStates}:
\begin{equation*}
    \left\{\ket{\mathsf{X}_{v}(s)} \ce \hat{U}_v(s )\ket{\mathsf{X_0}} \mbox{ for all } s\in[s_0,s_1]\subset\mathbb{R} \right\},
\end{equation*}
where
\begin{equation*}
    \hat{U}_v(s) \ce  e^{-is \hat{\mathsf{\Pi}}[v]} = e^{-is \int d\Sigma(x) \, v^{\mu}(x) \hat{\Pi}_\mu(x)}
\end{equation*}
is a unitary representation of  $G_v$. Now consider that
\begin{align}
    \hat{\mathsf{X}}[w] \ket{ \mathsf{X}_v(s)]} &= \hat{\mathsf{X}}[w] e^{-is \hat{\mathsf{\Pi}}[v]}\ket{\mathsf{X}_0} = \sum_{k=0}^\infty \frac{(-is)^k}{k!} \hat{\mathsf{X}}[w]  \hat{\mathsf{\Pi}}[v]^k\ket{\mathsf{X}}.
    \label{step1}
\end{align}
To simplify the sum, observe that
\begin{align*}
    \left[ \hat{\mathsf{X}}[w], \hat{\mathsf{\Pi}}[v]^2\right] &=\left[ \hat{\mathsf{X}}[w], \hat{\mathsf{\Pi}}[v]\right] \hat{\mathsf{\Pi}}[v] + \hat{\mathsf{\Pi}}[v] \left[ \hat{\mathsf{X}}[w], \hat{\mathsf{\Pi}}[v]\right]   = 2i(w, v) \hat{\mathsf{\Pi}}[v].
\end{align*}
Suppose that this holds for the $m$th case, so that
\begin{align*}
    \left[ \hat{\mathsf{X}}[w], \hat{\mathsf{\Pi}}[v]^m\right] = mi(w,v) \hat{\mathsf{\Pi}}[v]^{m-1}.
\end{align*}
Then it also holds for the $m+1$th case:
\begin{align*}
    \left[ \hat{\mathsf{X}}[w], \hat{\mathsf{\Pi}}[v]^{m+1}\right] 
    &= \left[ \hat{\mathsf{X}}[w], \hat{\mathsf{\Pi}}[v]^{m}\right] \hat{\mathsf{\Pi}}[v]  + \hat{\mathsf{\Pi}}[v]^{m} \left[ \hat{\mathsf{X}}[w], \hat{\mathsf{\Pi}}[v]\right] = (m+1)i(w,v)\hat{\mathsf{\Pi}}[v]^m.
\end{align*}
By induction it follows that $[ \hat{\mathsf{X}}[w], \hat{\mathsf{\Pi}}[v]^k ] = k i (w, v)\hat{\mathsf{\Pi}}[v]^{k-1}$, which allows for a further simplification of Eq.~\eqref{step1}:
\begin{align*}
    \hat{\mathsf{X}}[w] \ket{ \mathsf{X}_{v}(s)} & = \sum_{k=0}^\infty \frac{(-is)^k}{k!} \Big(  \hat{\mathsf{\Pi}}[v]^k \hat{\mathsf{X}}[w]  + k i (w, v)\hat{\mathsf{\Pi}}[v]^{k-1}  \Big) \ket{\mathsf{X}_0}  \nn \\
    & =  \Big(e^{-is\hat{\mathsf{\Pi}}[v]} \hat{\mathsf{X}}[w]  + i (w, v)  \sum_{k=1}^\infty \frac{(-is)^k}{(k-1)!}  \hat{\mathsf{\Pi}}[v]^{k-1} \Big) \ket{\mathsf{X}_0}  \nn\\
    & =  \Big( e^{-is\hat{\mathsf{\Pi}}[v]} {\mathsf{X}}_0[w]  +  i (w, v)  \sum_{m=0}^\infty \frac{(-is)^{m+1}}{m!}  \hat{\mathsf{\Pi}}[v]^{m} \Big) \ket{\mathsf{X}_0}  \nn\\
     & =  \left(   \mathsf{X}_0[w] e^{-is\hat{\mathsf{\Pi}}[v]} + (w, v) s e^{-is\hat{\mathsf{\Pi}}[v]}  \right) \ket{\mathsf{X}_0}  \nn\\
     & =  \big( \mathsf{X}_0[w] +  (w, v)s     \big) \ket{\mathsf{X}_{g_s^v}}  \nn\\
     & =  \int d\Sigma(x) \, w_\mu(x) \big(   X_0^\mu(x) + s v^\mu(x)   \big) \ket{\mathsf{X}_{g_s^v}}.
    \label{step2}
\end{align*}
The above equality must hold for all $w_\mu(x)$, which implies
\begin{align*}
    \hat{X}^{\mu}(x) \ket{\mathsf{X}_{v}(s)}  &= \Big(   X_0^\mu(x) + sv^\mu(x) \Big) \ket{\mathsf{X}_{v}(s)} ,
\end{align*}
as stated in Eq.~\eqref{coherntEigenTry2}.

\subsection{Commutation of $\hat{F}_A[\mathsf{X}_q]$ with the smeared constraint operator}
\label{DiracObservablesApeendix}

We can observe that $\hat{F}_{A}[\mathsf{X}_q]$ formally commutes with the constraint:
\begin{align*}
    \big[ \hat{\mathsf{C}}[g], \hat{F}_{A}[\mathsf{X}_q] \big] \ket{\Psi_{\rm phys}} &=  \hat{\mathsf{C}}[g]  \hat{F}_{A}[\mathsf{X}_q]  \ket{\Psi_{\rm phys}}  \\
    &= \hat{\mathsf{C}}[g]  \int \mathcal{D}q \, \hat{U}_{\mathsf{X}}[\mathsf{X}_{\bar{q}} - \mathsf{X}_{q}] \otimes \hat{U}_{\phi}[\mathsf{X}_{\bar{q}} - \mathsf{X}_{q}] \ket{\mathsf{X}_q} \! \bra{\mathsf{X}_q} \otimes\hat{A} \ket{\Psi_{\rm phys}}  \\
    &= \hat{\mathsf{C}}[g]  \int \mathcal{D}q \,\hat{U}_{\mathsf{X}}[\mathsf{X}_{\bar{q}} - \mathsf{X}_{q}] \otimes \hat{U}_{\phi}[\mathsf{X}_{\bar{q}} - \mathsf{X}_{q}]\ket{\mathsf{X}_q} \otimes \hat{A} \ket{\psi_{\phi}[\mathsf{X}_q]}.
\end{align*}
Let us now expand the conditional state in the eigenbasis of $\hat{A}$, so that  $\ket{\psi_{\phi}[\mathsf{X}_q]} = \sum_k \psi_k[\mathsf{X}_q] \ket{a_k}$, where $\hat{A} \ket{a_k} = a_k \ket{a_k}$ and $\psi_k[\mathsf{X}_q] \ce \braket{ a_k|\psi_{\phi}[\mathsf{X}_q]}$.
This allows us to express the above commutator as
\begin{align*}
    \big[ \hat{\mathsf{C}}[g], \hat{F}_{A}[\mathsf{X}_q] \big] \ket{\Psi_{\rm phys}}  &= \hat{\mathsf{C}}[g]  \int \mathcal{D}q \, \hat{U}_{\mathsf{X}}[\mathsf{X}_{\bar{q}} - \mathsf{X}_{q}] \otimes \hat{U}_{\phi}[\mathsf{X}_{\bar{q}} - \mathsf{X}_{q}] \ket{\mathsf{X}_q} \otimes     \hat{A}  \sum_k \psi_k \ket{a_k[\mathsf{X}_q]} \\
    &= \hat{\mathsf{C}}[g]  \int \mathcal{D}q \, \hat{U}_{\mathsf{X}}[\mathsf{X}_{\bar{q}} - \mathsf{X}_{q}] \otimes \hat{U}_{\phi}[\mathsf{X}_{\bar{q}} - \mathsf{X}_{q}] \ket{\mathsf{X}_q} \otimes \sum_k \psi_k a_k\ket{a_k[\mathsf{X}_q]} \\
    &= \sum_k \psi_k a_k \hat{\mathsf{C}}[g]  \left( \int \mathcal{D}q \, \hat{U}_{\mathsf{X}}[\mathsf{X}_{\bar{q}} - \mathsf{X}_{q}] \otimes \hat{U}_{\phi}[\mathsf{X}_{\bar{q}} - \mathsf{X}_{q}] \ket{\mathsf{X}_q} \otimes I_\phi \right)      \ket{a_k[\mathsf{X}_q]} \\
    &= \sum_k \psi_k a_k \hat{\mathsf{C}}[g]  \mathcal{R}^{-1}_{\bf S}[\mathsf{X}_q] \ket{a_k[\mathsf{X}_q]} \\
    &= \sum_k \psi_k a_k \hat{\mathsf{C}}[g]  \ket{\bar{\Psi}_{{\rm phys},k}} \\
    &=0,
\end{align*}
where going from the third to fourth equality we have made use of the definition of the inverse of the reduction map given in Eq.~\eqref{defReductionInverse}.

\subsection{Heisenberg equation from deparametrization}

Starting with a Heisenberg picture of the form
\begin{equation*}
     \hat{A}_\phi(q) \ce \mathcal{R}_{\mathbf H}  \hat{F}_{A_0}[\mathsf{X}_q] \mathcal{R}_{\mathbf H}^{-1} = \hat{U}_{\phi}[\mathsf{X}_{0}  - \mathsf{X}_{q}]\hat{A}_0\,\hat{U}^\dagger_{\phi}[\mathsf{X}_{0}-\mathsf{X}_{q}],
\end{equation*}
where, as defined in the text: 
\begin{equation*}
    \hat{U}_{\phi}[\mathsf{X}_{0} - \mathsf{X}_{q}]\ce e^{ -i \!\int\! d\Sigma(x) \left(X_0^\mu(x) - X_q^\mu(x) \right) \hat{h}_{\mu}[X_q^\mu(x)]}.
\end{equation*}

Taking the functional derivative of the Heisenberg-picture operator:
\begin{align*}
    \int d\Sigma(y) \frac{\delta \hat{A}_\phi[\mathsf{X}_q]}{\delta X^\mu_q (y)}&=\int d\Sigma(y) \frac{\delta }{\delta X^\mu_q (y)}\left(e^{ -i \!\int\! d\Sigma(x) \left(X_0^\nu(x) - X_q^\nu(x) \right) \hat{h}_{\nu}[\mathsf{X}_q,x]} \hat{A}_0e^{ i \!\int\! d\Sigma(x) \left(X_0^\nu(x) - X_q^\nu(x) \right) \hat{h}_{\nu}[\mathsf{X}_q,x]}\right)\\
    &=\int d\Sigma(x)d\Sigma(y)\left[i\delta(x-y)\hat{h}_\mu[\mathsf{X}_q,x]-i  \left(X_0^\nu(x) - X_q^\nu(x) \right)\frac{\delta \hat{h}_\nu[\mathsf{X}_q,x]}{\delta X^\mu_q(y)} \right]\hat{A}_\phi[\mathsf{X}_q] \\
    &\quad +\int d\Sigma(x)d\Sigma(y)\hat{A}_\phi[\mathsf{X}_q]\left[-i\delta(x-y)\hat{h}_\mu[X_q,x]+i  \left(X_0^\nu(x) - X_q^\nu(x) \right)\frac{\delta \hat{h}_\nu[\mathsf{X}_q,x]}{\delta X^\mu_q(y)} \right] \\
    &=i \left[ \hat{H}_\mu[\mathsf{X}_q], \hat{A}_\phi[\mathsf{X}_q]\right] -  i\int d\Sigma(x)d\Sigma(y) \left(X_0^\nu(x) - X_q^\nu(x) \right) \left[ \frac{\delta \hat{h}_\nu[\mathsf{X}_q,x]}{\delta X^{\mu}(y)}, \hat{A}_\phi[\mathsf{X}_q]\right]\\
    &=i \left[ \hat{H}_\mu[\mathsf{X}_q], \hat{A}_\phi[\mathsf{X}_q]\right] -  i\int d\Sigma(x)d\Sigma(y) \left(X_0^\nu(x) - X_q^\nu(x) \right) \left[ \frac{\delta n_\rho(x) }{\delta X^{\mu}(y)} {T^\rho}_\nu, \hat{A}_\phi[\mathsf{X}_q]\right] \\
    &=i \left[ \hat{H}_\mu[\mathsf{X}_q], \hat{A}_\phi[\mathsf{X}_q]\right] -  i\int d\Sigma(x)d\Sigma(y) \left(X_0^\nu(x) - X_q^\nu(x) \right) \left[  \epsilon_{\rho \sigma} \left( \partial_x \frac{\delta  X^{\sigma}(x)}{\delta X^{\mu}(y)} \right) {T^\rho}_\nu, \hat{A}_\phi[\mathsf{X}_q]\right] \\
    &=i \left[ \hat{H}_\mu[\mathsf{X}_q], \hat{A}_\phi[\mathsf{X}_q]\right] -  i\int d\Sigma(x)d\Sigma(y) \left(X_0^\nu(x) - X_q^\nu(x) \right) \left[  \epsilon_{\rho \mu} \left( \partial_x  \delta(x-y) \right) {{\hat{T}}^\rho}_{\ \nu}(x), \hat{A}_\phi[\mathsf{X}_q]\right] \\
    &=i \left[ \hat{H}_\mu[\mathsf{X}_q], \hat{A}_\phi[\mathsf{X}_q]\right] -  i\int dxd\Sigma(y) \delta(x-y)   \left[  \epsilon_{\rho \mu}  \partial_x \left(\sqrt{\gamma(x)} \left(X_q^\nu(x) - X_0^\nu(x) \right)  {{\hat{T}}^\rho}_{\ \nu}(x) \right), \hat{A}_\phi[\mathsf{X}_q]\right] \\
    &=i \left[ \hat{H}_\mu[\mathsf{X}_q], \hat{A}_\phi[\mathsf{X}_q]\right] -  i  \left[  \int dx \,  \epsilon_{\rho \mu}  \partial_x \left(\sqrt{\gamma(x)} \left(X_q^\nu(x) - X_0^\nu(x) \right)  {{\hat{T}}^\rho}_{\ \nu}(x) \right), \hat{A}_\phi[\mathsf{X}_q]\right] \\
    &=i \left[ \hat{H}_\mu[\mathsf{X}_q], \hat{A}_\phi[\mathsf{X}_q]\right],
\end{align*}
where the integral appearing in the penultimate equality vanishes given that it is equal to a boundary term that vanishes under the assumption that the matter field stress-energy tensor vanishes at infinity. We thus recover the Heisenberg equation as stated in Eq.~\eqref{Hisenberg}.

\section{Foliation independence through equivalence of evolutions of the conditioned state}
\label{app: unitaryeqiv}

We give here the derivation of Eq.~\eqref{FoliationIndependence}, which illustrates the foliation independence of the evolution of the field degrees of freedom from an initial to a final hypersurface. The unitary operator evolves the conditioned state degrees of freedom through the integrated Schr\"odinger equation. It is also equivalent to mapping back the conditioned state to the perspective-neutral physical Hilbert space and then projecting back to the final embedding through a composition of the Schr\"odinger reduction maps. We start with the general map:
\begin{equation*}
\label{eq: evol_through_perpneutr}
    \hat{U}_{\phi}[\mathsf{X}_{q'} - \mathsf{X}_{q}] \ce e^{-i \int d\Sigma(x) \, \left( X^{\mu}_{q'}(x) - X^{\mu}_{q}(x) \right) h_\mu(x) }. 
\end{equation*}
We will now show that given a foliation that connects two embeddings, mapping from the initial embedding to the perspective-neutral space and back into the final embedding is equivalent to projecting the field on the initial embedding and then evolving it through an arbitrary foliation using the Hamiltonian flux. This is only possible due to the fact that physical states are zero eigenvalue states of the constraint and are hence gauge-invariant.

Consider a discrete one-parameter family $\mathcal{F} \ce \{ \mathsf{X}(t),  \forall t \in [t_i,t_f] \}$ consisting of $n+1$ spacelike hypersurfaces $\mathsf{X}(t)$ ``foliating'' Minkowski space such that $\mathsf{X}(t_i)=\mathsf{X}_q$ and $\mathsf{X}(t_f)=\mathsf{X}_{q'}$. This foliation is highly non-unique and each surface is separated from the previous and successive one by a small finite interval $\Delta t$ of the time coordinate $t$ such that: 
\begin{equation*}
    \Delta t=\frac{t_f - t_i}{n+1}
\end{equation*}
and $t_{k+1}=t_k+\Delta t$. Let us now consider the unitary evolution operator of the conditioned state between the $k$-th and the $(k+1)$-th surfaces of the foliation: 
\begin{equation*}
    U_{\phi}[\mathsf{X}(t_k+\Delta t) - \mathsf{X}(t_k)] = e^{-i \int d\Sigma(t_k,x) \, \Delta t\left( \frac{X^{\mu}(t_k+\Delta t,x) - X^{\mu}(t_k,x)}{\Delta t} \right) h_\mu(t_k,x) }. 
\end{equation*}
The evolution between the initial and final leaves of the foliation is now obtained by applying the unitary evolution between two adjacent slices $n$ amount of times:
    \begin{align*}
        \hat{U}_{\phi}[\mathsf{X}(t_f) - \mathsf{X}(t_i)]&=\mathcal{P}\prod_{k=1}^{n}\hat{U}_{\phi}[\mathsf{X}(t_k+\Delta t) - \mathsf{X}(t_k)] \nn\\     &=\mathcal{P}\prod_{k=1}^{n} \exp\left[-i \int d\Sigma(t_k,x) \,\Delta t \left( \frac{X^{\mu}(t_{k+1},x) - X^{\mu}(t_k,x)}{\Delta t} \right) \hat{h}_\mu(t_k,x) \right] \nn \\
        &=\mathcal{P} \exp\left[-i \sum_{k=1}^{n}\Delta t\int d\Sigma(t_k,x) \, \left( \frac{X^{\mu}(t_{k+1},x) - X^{\mu}(t_k,x)}{\Delta t} \right) \hat{h}_\mu(t_k,x) \right].
    \end{align*}
The path-ordering operator ensures that the unitary evolution acts in the correct order on the leaves of the foliation, and allows us not to worry about whether the Hamiltonian flux commutes at different points of the foliation. We now take the limit for an infinite number of infinitesimally close slices, we recover Eq.~\eqref{evolgen}:
    \begin{align} 
        \hat{U}(t_f,t_i)&=\lim_{       \substack{\Delta t\to 0\\n\rightarrow\infty}}\hat{U}_{\phi}[\mathsf{X}(t_f) - \mathsf{X}(t_i)]\nn \\
        &=\lim_{\substack{\Delta t\to 0\\n\rightarrow\infty}}\mathcal{P} \exp\left[-i \sum_{k=1}^{n}\Delta t\int d\Sigma(t_k,x) \, \left( \frac{X^{\mu}(t_{k+1},x) - X^{\mu}(t_k,x)}{\Delta t} \right) \hat{h}_\mu(t_k,x) \right] \nn\\
        &=\mathcal{P} \exp\left(-i \int_{\mathsf{X}(t_i)}^{\mathsf{X}(t_f)} dt d\Sigma(t,x) \, t^\mu(t,x) \hat{h}_\mu(t,x) \right),
        \label{eq: foliationevolution}
    \end{align}
where $t^\mu(t,x) \ce\partial_t X^\mu=n^\mu(t,x) N(t,x)+N^\mu(t,x)$ is the deformation vector field. 

Alternatively, before taking the continuum limit, the recovery map and its inverse, defined in Eqs.~\eqref{defReduction} and \eqref{defReductionInverse}, can be used to map between successive leaves:
 \begin{align*}
        \hat{U}_{\phi}[\mathsf{X}(t_f) - \mathsf{X}(t_i)]&=\mathcal{P}\prod_{k=1}^{n}\hat{U}_{\phi}[\mathsf{X}(t_k+\Delta t) - \mathsf{X}(t_k)] \nn\\
        &=\mathcal{P}\prod_{k=1}^{n}  \mathcal{R}_{\mathbf S}[\mathsf{X}(t_{k+1})]  \circ \mathcal{R}_{\mathbf S}^{-1}[\mathsf{X}(t_k)] \\
        &= \mathcal{R}_{\mathbf S}[\mathsf{X}(t_f)] \left( \mathcal{P}\prod_{k=1}^{n-1}  \mathcal{R}_{\mathbf S}^{-1}[\mathsf{X}(t_{k+1})]  \circ \mathcal{R}_{\mathbf S}[\mathsf{X}(t_k)] \right) \mathcal{R}_{\mathbf S}^{-1}[\mathsf{X}(t_i)] \\
        &= \mathcal{R}_{\mathbf S}[\mathsf{X}(t_f)] \circ \mathcal{R}_{\mathbf S}^{-1}[\mathsf{X}(t_i)],
    \end{align*}
where in moving from the third line, we have used the fact that $\mathcal{R}_{\mathbf S}^{-1}[\mathsf{X}(t_{k+1})]  \circ \mathcal{R}_{\mathbf S}[\mathsf{X}(t_k)] = \hat{I}_\phi$. The final expression depends only on the initial and final hypersurfaces, not discrete foliation. Thus the limit $\Delta t \to 0, \, n\to\infty$ can be taken with the result that 
\begin{equation}
    \hat{U}(t_f,t_i) = \mathcal{R}_{\mathbf S}[\mathsf{X}(t_f)] \circ \mathcal{R}_{\mathbf S}^{-1}[\mathsf{X}(t_i)].
    \label{eq: foliationevolution2}
\end{equation}
Given that Eqs.~\eqref{eq: foliationevolution} and \eqref{eq: foliationevolution2} must be equal, it follows that the evolution between $\mathsf{X}(t_i)$ and $\mathsf{X}(t_f)$ is independent of the choice of foliation, despite the foliation dependence of the integrand in Eq.~\eqref{eq: foliationevolution}.

\section{Discussion of the anomaly}
\label{anomaly}
In this paper, we have considered the quantization of a scalar field on a two-dimensional spacetime. In the massless case, this constitutes a CFT. Hence, we need to address the conformal anomaly arising in the process. Since the conformal group is a subgroup of the spacelike preserving diffeomorphisms that evolve the embeddings, it carries the Virasoro algebra into the Hamiltonian flux when the constraint $C_\mu(x)$ is smeared by a conformal vector. Following Kucha\v{r} \cite{kuchar_dirac_1989}, we see how the anomaly arising from the commutation relation of the Hamiltonian flux can be expressed in terms of derivatives of an anomaly potential $A_\mu$. This potential needs to be incorporated with the Hamiltonian flux, such that the right commutation relations will be ensured. Following Kucha\v{r}'s notation, the anomaly potential can be spilt into two parts:
    \begin{equation*}
	    A_\mu(x)= {_I}A_\mu(x) + {_{III}}A_\mu(x).
	\end{equation*}
The first term is related to a soft breaking of the conformal symmetry by the introduction of a macroscopic length scale. For example, in Kucha\v{r}, it is given for the choice of manifold being $\mathbb{R}\times S^1$. If we assume the stress-energy tensor of the field to be normally ordered such that its vacuum expectation value on the plane vanishes, we can interpret the first part of the anomaly potential as a Casimir energy given by the periodic boundary condition \cite{kuchar_dirac_1989}. It is embedding-independent and vanishes for the unbounded $\mathbb{R}^2$ manifold considered in this paper.
    \begin{equation*}
        {_I}A_\mu(x)=0
    \end{equation*}
The second part of the anomaly is instead embedding-dependent. It relates the distribution of the energy along the embedding to the trace of its extrinsic curvature $\mathcal{K}(x)={\mathcal{K}^\mu}_\mu(x)$:
    \begin{equation*}
        {_{III}}A_\mu(x)=\sqrt{\gamma(x)}\left(X_\mu(x)\frac{\partial\mathcal{K}(x)}{\partial x}-\mathcal{K}(x)^2n_\mu(x)\right).
    \end{equation*}
As explained in Sec.~\ref{sec: PFT}, this vanishes for embeddings with zero extrinsic curvature and when integrated over asymptotically flat embeddings. Given a spacetime embedding $X^\mu(t,x)=(T(t,x),X(t,x))$. We can define the oriented Minkowski basis vectors as $t_\mu=-\partial_\mu T$, $s_\mu=\partial_\mu X$ such that $t_\mu t^\mu=-1$, $s_\mu s^\mu=1$ and $t^\mu s_\mu=0$.
Following \cite{kuchar_dirac_1989}, we can rewrite this part of the anomaly such that when integrated
    \begin{align*}
        &\int d\Sigma(x) \, {_{III}}A_\mu(x)t^\mu[X^\mu(x)]=-\int d\Sigma(x) \, \partial_x(\sqrt{\gamma}\mathcal{K}(x)t^0)=0, \\
        &\int d\Sigma(x)\, {_{III}}A_\mu(x)s^\mu[X^\mu(x)]=-\int d\Sigma(x) \, \partial_x(\sqrt{\gamma}\mathcal{K}(x)s^1)=0,
    \end{align*}
and we see that the anomaly vanishes as the extrinsic curvature vanishes at infinity.

\twocolumngrid

\bibliography{PFT}

\end{document}